# Local electronic transport across probe/ionic conductor interface in scanning probe microscopy

K.N. Romanuk[1,2], D.O. Alikin[1], B.N. Slautin[1], A. Tselev[2], V.Ya. Shur[1] and A.L. Kholkin[2]

[1] School of Natural Sciences and Mathematics, Ural Federal University, Ekaterinburg, Russia
[2] Department of Physics and CICECO – Aveiro Institute of Materials, University of Aveiro, 3810-193 Aveiro, Portugal

## Abstract

Charge carrier transport through the probe-sample junction can have substantial consequences for outcomes of electrical and electromechanical atomic-force-microscopy (AFM) measurements. For understanding physical processes under the probe, we carried out conductive-AFM (C-AFM) measurements of local current-voltage (I-V) curves as well as their derivatives on samples of a mixed ionic-electronic conductor $Li_{1-x}Mn_2O_4$ and developed an analytical framework for the data analysis. The implemented approach discriminates between contributions of a highly resistive sample surface layer and bulk with account of ion redistribution in the field of the probe. It was found that with increasing probe voltage, the conductance mechanism in the surface layer transforms from Pool-Frenkel to space-charge-limited current. The surface layer significantly alters the ion dynamics in the sample bulk under the probe, which leads, in particular, to a decrease of the effective electromechanical AFM signal associated with the ionic motion in the sample. The framework can be applied for the analysis of mechanisms of electronic transport across the probe/sample interface as well as the role of the charge transport in the electric field distribution, mechanical, and other responses in AFM measurements of a broad spectrum of conducting materials.

## Introduction

Scanning probe microscopy finds a growing variety of applications for probing local electronic and ionic transport in the vicinity of the different kinds of interfaces, such as active surfaces in ionic conductors [1–9], battery separators [10,11], metal-oxide interfaces in resistive switching elements [12], layers in ferroelectric hetero-structures [13,14], and ferroelectric domains and phase boundaries [15]. Importantly, in contrast to macroscopic measurements, tip-induced voltage-driven responses take place in a highly inhomogeneous electric field and, thus, involve different phenomena, such as local electrostatic interactions [16–18], charge injection [19–21], Joule heating [22], field-induced phase transformation [23,24], and others. Quantitative analysis of the surface displacements induced by an electrically biased probe in contact with a material in a number of imaging modes including electrostatic force microscopy, piezoresponse force microscopy, or electrochemical strain microscopy (ESM) needs deeper understanding of electronic transport across the probe/sample interface [25–28]. This involves elucidating the role of surface and thin surface layers, which may differ in structure and properties from bulk, in charge transport, electric field distribution, mechanical and other responses.

For example, in ESM measurements, detailed consideration of the charge transport through the probe/surface interface between a solid ionic conductor (SIC) or a mixed ionic-electronic conductor (MIEC) is required for correct evaluations of ion concentration and mobility distributions. As was found in [25], local diffusion coefficients and ionic concentrations can be extracted from frequency dependences of the ESM response using an electron-conducting probe. The presence of the electronic current across the platinum-coated probe/$LiMn_2O_4$ interface leads to a $\sim 1/f^{1/3}$ dependence of the ESM response with a value ≥1 pm/V in the frequency range from 3 to 30 kHz [25]. Along with a sufficiently high magnitude of the response, a rigorous analysis of electronic transport processes across the probe/sample interface is implicitly relevant for the proper understanding of the physical processes under the probe and for the correct estimation of the concentration and ion mobility distributions from ESM signals.

$Li_{1-x}Mn_2O_4$ (LMO) spinel is a mixed-valence ($Mn^{3+}$/$Mn^{4+}$) MIEC compound with a total conductivity dominated by the electronic contribution. Its electronic conduction is due to electron-hopping between high-valence ($Mn^{4+}$) and low valence ($Mn^{3+}$) cations [29]. Conductivity of this type is known as the polaron hopping conduction. In LMO, it is due to hole ($\eta_+$) and electron ($\eta_-$) polarons [30]. From the infrared (IR) spectroscopy data, LMO can be considered as a donor-doped ($Li^+$), n-type semiconductor [31]. The electron density, $n_{e^-}$, increases with increasing Li concentration, $n_{Li}$, because Li donates its electron to the lattice and becomes a $Li^+$-ion [30], which is coupled with oxidation of $Mn^{3+}$ to $Mn^{4+}$ [32,33]. On the other hand, the Mn-Mn distance increases with lithiation, and it can result in an increase of the conductance activation energy[34–36]. The electron conductivity, $\sigma(n_{Li})$, as a function of Li concentration can be influenced by these two opposite effects. According to the literature [32–37] , there is an evident conductivity dependence in $Li_{1-x}Mn_2O_4$ on Li content within a range $0 < x < 1$. Therefore, I-V curve mapping can be used to obtain the information about local ion concentration distribution near the surface.

In local measurements with an AFM probe, the electric field distribution in the sample in the vicinity of the point-like tip-sample contact is significantly nonuniform and peaks under the tip apex. The local resistivity under the tip is the main contribution to the full impedance of the probe-sample system. Apparently, it is defined by the local electric conduction of the sample material and directly linked to the electron/ionic concentration. From this point of view, one may expect an asymmetry of I-V curves relative to the polarity of the bias voltage applied between the AFM probe and LMO sample [7,25]. The asymmetry, even in the ohmic conduction regime, can be significant being associated with the Li-ion redistribution under the probe. Furthermore, due to the presence of surface and bulk charge traps, conducting properties of the material are nonuniform. Compared to the bulk, a much higher trap density is expected at and near the sample surface. For example, charge traps may be associated with dangling bonds and electronic surface states whose energy levels are deep inside the semiconductor band gap [38]. I-V curves measured on nominally identical samples can exhibit a large spread in their behavior because of different surface states [39]. The trap influence is manifested by dependence of local conductivity on local trap density through different possible conduction mechanisms, such as Space-Charge-Limited Currents (SCLC) [40–42] and Poole–Frenkel conduction, which can be dominating at the small radius of the probe apex (~30 nm) and under the high electric field near the probe [43,44]. The conductivity behavior can be further complicated by the effect of voltage-dependent contributions of surface and bulk traps to the charge transport.

In the present work, we have performed a comprehensive analysis of the I-V curves measured on the surface of active ionic particles in the cross-sections of commercial LMO battery cathodes. Using experimental data and rigorous calculations, we established basic mechanisms of charge transport across the probe/LMO interface. The obtained results are of direct importance for evaluation of electromechanical responses measured with SPM in MIECs. In turn, possible applications of the developed analytical framework are not limited to MIECs, and it can be implemented for a broad range of electrically conducting materials.

In the first part of the paper, we consider possible charge transport mechanisms occurring in the surface layer at different applied bias voltages. The voltage ranges for the Poole–Frenkel conduction and for the SCLC regime are established based on experimental I-V curves. An analytical expression for the Poole–Frenkel conduction for a point contact in a spherical approximation is obtained and used for the analysis of the experimental data to estimate the thickness and conductivity of the surface layer.

In the second part, we analyze and interpret the experimental I-V curves using an equivalent electric circuit model. Values of the surface layer and bulk resistivities are considered as power functions of the current. The obtained analytical expressions describe well the experimental I-V dependences. The extracted fitting parameters are used to establish the voltage ranges for ohmic and SCLC conduction regimes in the bulk and for evaluation of the bulk resistivity contribution into the full impedance of the system.

Finally, we consider possible Li-ion redistribution responsible for the changes in the bulk conductivity at different probe voltage polarities. In particular, we obtain relations for the steady state bulk resistivity as a function of the probe current. The analysis of the experimental I-V curves shows that the values of Li-ion concentrations outside of the space charge region under the probe differ 1.5–2 times for the different polarities of the applied bias voltage. Together with SCLC mechanism of the charge transport, it results in asymmetry of the I-V curves.

## Experimental details and methods

C-AFM measurements were performed on Li-battery positive electrodes (cathodes) comprising LMO ceramic particles embedded in an PVDF/carbon matrix [23]. The samples were provided by Robert Bosch, GmbH. The cathodes were extracted from fresh commercial cells in the fully charged state (the lattice parameter of the cubic spinel $Li_{0.61}Mn_2O_4$ samples was 0.809 nm), washed in dimethyl carbonate, embedded in epoxy resin and rigorously polished with step-by-step decreasing abrasive (up to ¼ um) and then with Ar ion beam milling. C-AFM measurements were performed with a NTEGRA Aura scanning probe microscope (NT-MDT, Russia) in vacuum at ~$10^{-3}$ Torr. The cathode samples were grounded through the metallic bottom electrode connected to an Al current collector. $W_2C$ coated cantilever probes (Scansens, Germany, HA_NC_W$_2$C+) with about 10 N/m spring constants and 130 kHz resonance frequencies were used for the measurements. An LMO particle in a homogeneous lithiation state was chosen based on data of confocal Raman spectroscopy (see Supplementary 1 for details). The particle topography was inspected with AFM in the tapping mode in order to reduce the probe wear due to the tip interaction with topographic features.

Local I-V characteristics were measured directly via registration of current under dc bias as well as with the use of a lock-in technique, when a superposition of dc and ac voltages $U_{ex}=U_{dc}+U_{ac}$ was applied to the AFM tip (Figure 1). In the latter case, ac voltage of an amplitude $U_{ac}$=1V and a frequency of $f$ =300 Hz was applied simultaneously to the sample and to the reference input of an internal lock-in amplifier. A low-pass filter with a cut-off frequency of 15 Hz was used to provide the most accurate, low-noise measurements. The lock-in amplitude signal was plotted as a function of the applied dc voltage. The phase signal changed negligibly. The dc current, $1^{st}$ and $2^{nd}$ harmonics signals were obtained simultaneously. The lock-in signals for the $1^{st}$ and $2^{nd}$ harmonics of the applied ac voltage are dominated by the first and second derivatives of the current over voltage, respectively, but also include additional undesirable contributions from higher derivatives. The higher-order contributions are decreasing with the derivative number and can be reduced by decreasing the magnitude of $U_{ac}$. In our measurements, the lock-in signals at the $1^{st}$ and $2^{nd}$ harmonics were interpreted as proportional to the first and second derivatives of the I-V curves. Numerical integration of signals of the $1^{st}$ and $2^{nd}$ harmonics was performed for recovering the initial I-V dependence to suppress the $1/f$ noise contribution and improve the signal-to-noise ratio. At each integration step, integration constants were calculated from the difference between the initial I-V and integrated $\frac{\partial I(U)}{\partial U}$ curves at 0 V. A background appearing as a result of integration was subtracted before application of the integration constants. Finally, the integrated I-V curves were scaled by the current values measured at dc in order to correspond to the real current values.

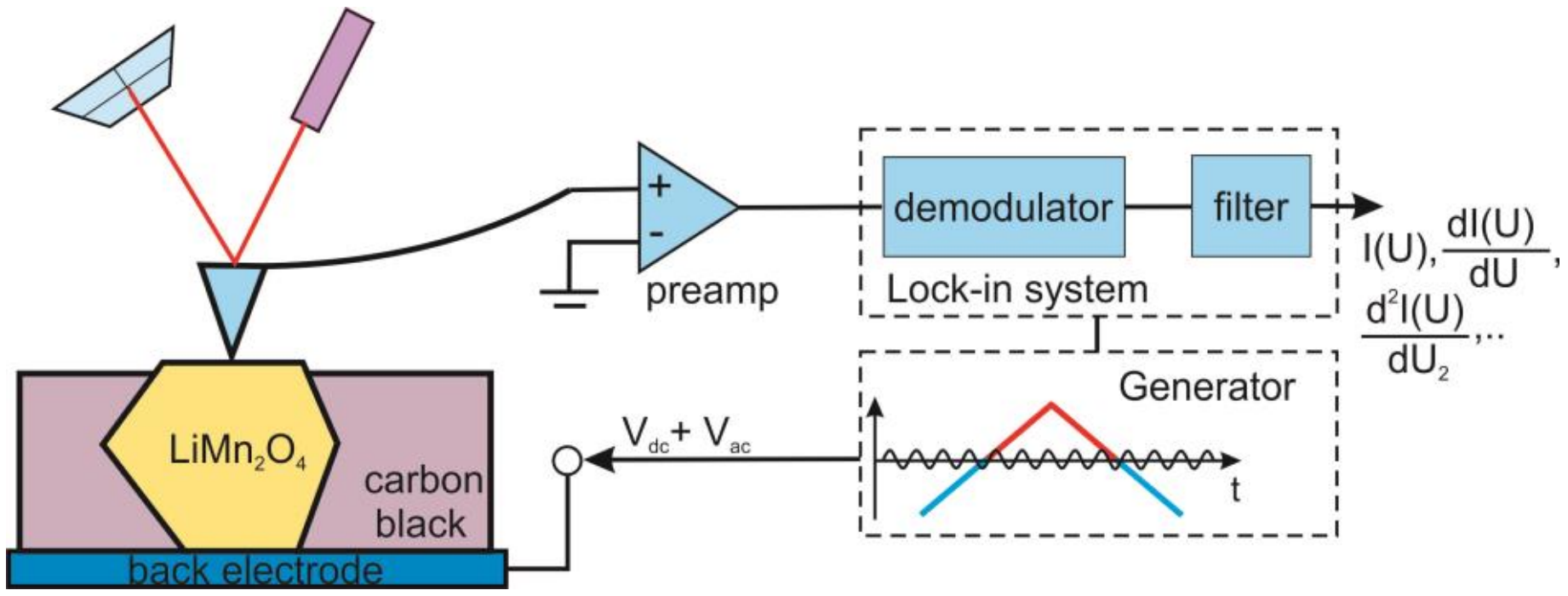

Figure 1. Schematic of the C-AFM measurements in the $LiMn_2O_4$ cathode. Bias voltages, ac and dc, were applied to the back electrode; probe is grounded.

The probe-sample loading force during the measurements was varied in a range between 20 nN and 1100 nN. All measurements and estimations were performed under an assumption that the local temperature under tip does not differ significantly from the ambient temperature. Calculations for estimation of the temperature in the tip-sample junction during measurements are presented in Supplementary 2. Kelvin probe force microscopy (KPFM) measurements were carried out with the same probes under low-vacuum ($10^{-3}$ Torr) in the double-pass mode. The vertical distance between the passes was chosen to be around 10 nm. An ac voltage of 0.1 V in amplitude was applied to the tip to capture the electrostatic response. Calibration of the work functions was made on a freshly cleaved pyrolytic graphite.

## Results and discussions

### 1. Charge transport mechanisms at the probe/LMO interface.

#### 1.1 *High bias voltage*

The I-V curves measured by the C-AFM on the surface of semiconducting materials can be associated with different charge transport processes across the probe-sample interface [40–42,45–51]. Figure 2 displays typical I-V curves obtained in our C-AFM experiments with the LMO samples. As illustrated in Figure 2, the experimental I-V curves can be best fit with a power-law function $I \sim V_{dc}^{\alpha}$ with $\alpha \geq 2$ for both bias voltage polarities at $|U_{dc}| > 1$ V. This indicates that the dominating charge-transport mechanism at biases $|U_{dc}| > 1$ V is SCLC in agreement with previously reported data for LMO [25].

To interpret this observation, we note that the AFM probe with a radius of 10-30 nm can serve as an effective charge injection electrode. A schematic of a metal/n-type semiconductor contact relevant for the electron injection in the sample is shown in Fig. 3. Due to a small radius of the probe tip apex (~30 nm) and a small width of the Schottky barrier at the probe-sample interface, which results from a high density of charge carriers—$Li^{+}$ ions and electrons (as $\eta_{-}$ polarons [30])—in the LMO, injection of both electrons and holes can take place through the metal/LMO junction. The width of the interfacial Schottky barrier in LMO can be estimated equal to the screening length in the LMO $\lambda_S$ ~0.3 nm (see Supplementary Information 3). The height of the interfacial Schottky barrier is $\phi_B \approx E_g - \phi_m + \chi$ and $\phi_B \approx \phi_m - \chi$ for electrons and holes, respectively (Fig. 3 (c) and (e)), where $E_g$ is the band gap, $\phi_m$ is the work function of the metal and $\chi$ is the electron affinity of the semiconductor; the charge injection occurs when the voltage applied to the junction $U < E_F - E_C$ for electrons and $U > E_F - E_V$ for holes.

Note that the alternative charge transport mechanism through the probe-sample interface can be the Richardson-Schottky injection (thermionic emission). It takes place when semiconductor work function is lower than metal work function. Therefore, to further support the interpretation of the experimental I-V characteristics, work function measurements were performed on the LMO particle.

The measurements were carried out using the Kelvin probe force microscopy (KPFM) mode with $W_2C$-coated probes in ambient. They showed a work function difference between the probe and LMO $\phi_{LMO} - \phi_{W_2C} =$ 0.3-0.5 eV. The probe work function $\phi_{W_2C} = 4.7$ eV was extracted from calibration measurements made on a surface of a freshly cleaved highly oriented pyrolytic graphite (HOPG). (See also Table S4 for work function values for different tip coating materials in comparison with the work function of LMO; for all available probes, the work functions are lower than that of LMO.) With this value, the estimated work function for LMO $\phi_{LMO} =$ 5.0-5.2 eV, which is ~1 eV lower than the experimental value found in the literature $\phi_{LMO} =$ 6.2 eV. Such a difference can be explained by the used surface preparation conditions. Specifically, the work function decrease can be induced by the ion-beam processing of the sample surface. A similar effect was observed for oxide materials after exposure to ultraviolet light [52]. Still, since the semiconductor work function is higher than the metal work function, the Richardson-Schottky injection can be excluded as a dominating charge transport mechanism in our C-AFM measurements. It should be noted that, independently of the preparation conditions, experiments reported in literature on different LMO samples showed a similar nonlinear (SCLC) character of C-AFM I-V curves [53,54].

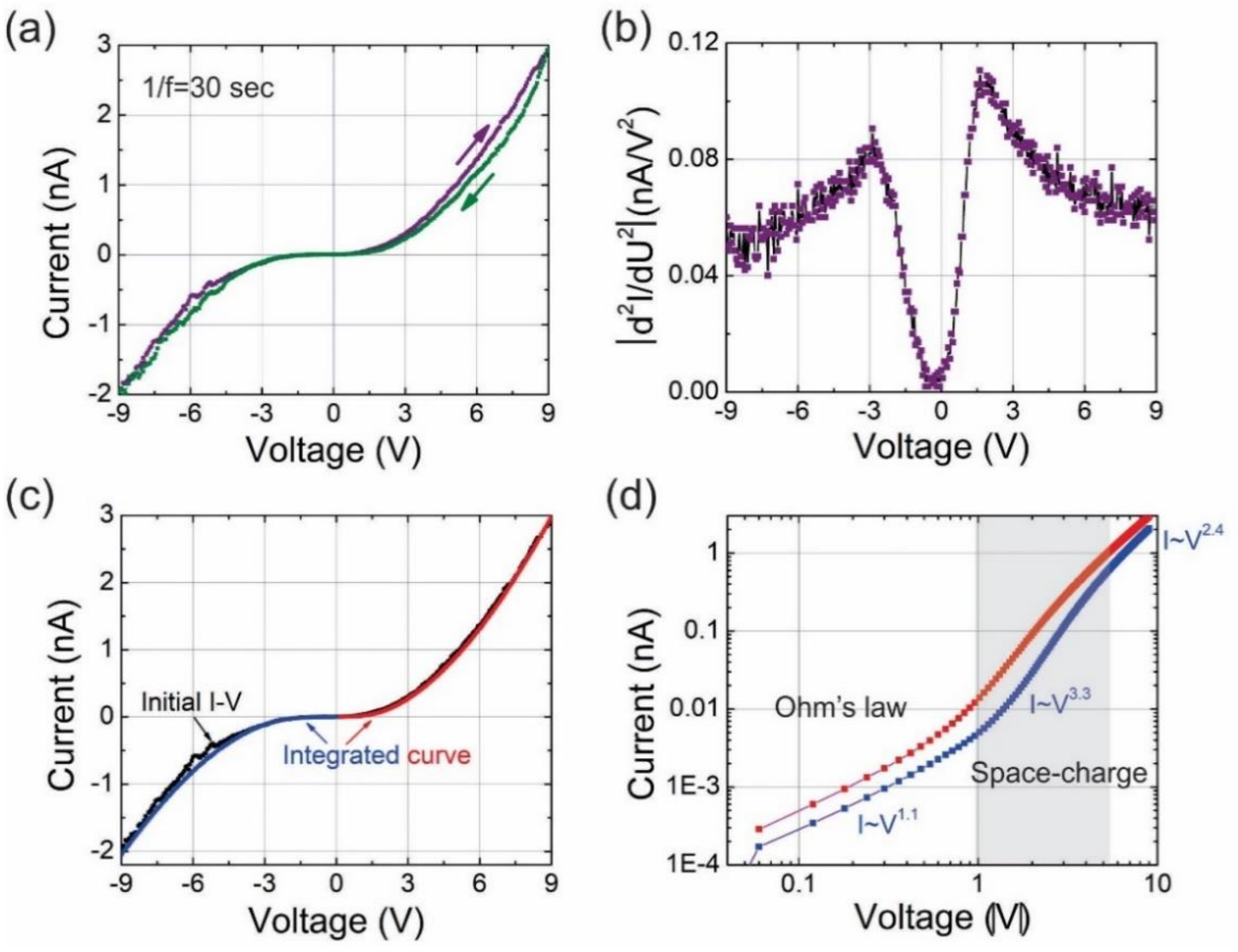

Figure. 2. I-V curves measured at a 880 nN probe-sample loading force for 30 s per curve. (a) I-V curves acquired between $U_{dc} = -9$ V and +9V in forward and backward directions. (b) Absolute value

of the second derivative, $\left|\frac{\partial I^2(U)}{\partial U^2}\right|$ acquired using the lock-in technique simultaneously with the I-V curves in (a). (c), (d) I-V curves obtained from the second derivative curves in (b) after double-integration over $U_{dc}$. The blue and red lines correspond respectively to the negative and positive bias voltages on the back electrode. (d) I-V curves of (c) shown in the log-log scale. Near-ohmic conduction was observed between -1 V and +1 V of the dc bias voltage.

### 1.2 *Low bias voltage*

At low enough bias voltages and electric field in the probe-sample junction, an near-ohmic conduction with a linear I-V dependence is possible based on the analysis of the energy level diagrams in Fig. 3. As can be inferred from the diagrams in Fig. 3, the near-ohmic conduction takes place at $E_F - E_C < U < E_F - E_V$ (Fig. 3(b) and (d)) involving charge transfer from the conduction band of the metal to the surface charge traps followed by the field-enhanced thermal excitation and detrapping of the trapped electrons or holes. In the case shown in Fig. 3(b), the charge flow occurs in part due to intrinsic electrons in LMO moving from the bottom of the LMO conduction band to the empty states above the Fermi level of the metal conduction band. Taking the width of the energy gap in LMO $E_g$ ~2.1 eV (see Table S5) and the Fermi level position at about 1 eV from the bottom of the LMO conduction band [30], the bias voltage interval for observation of the near-ohmic conduction is -1 V $< U <$ 1 V, in agreement with the experimentally obtained I-V curves in Fig. 2(d).

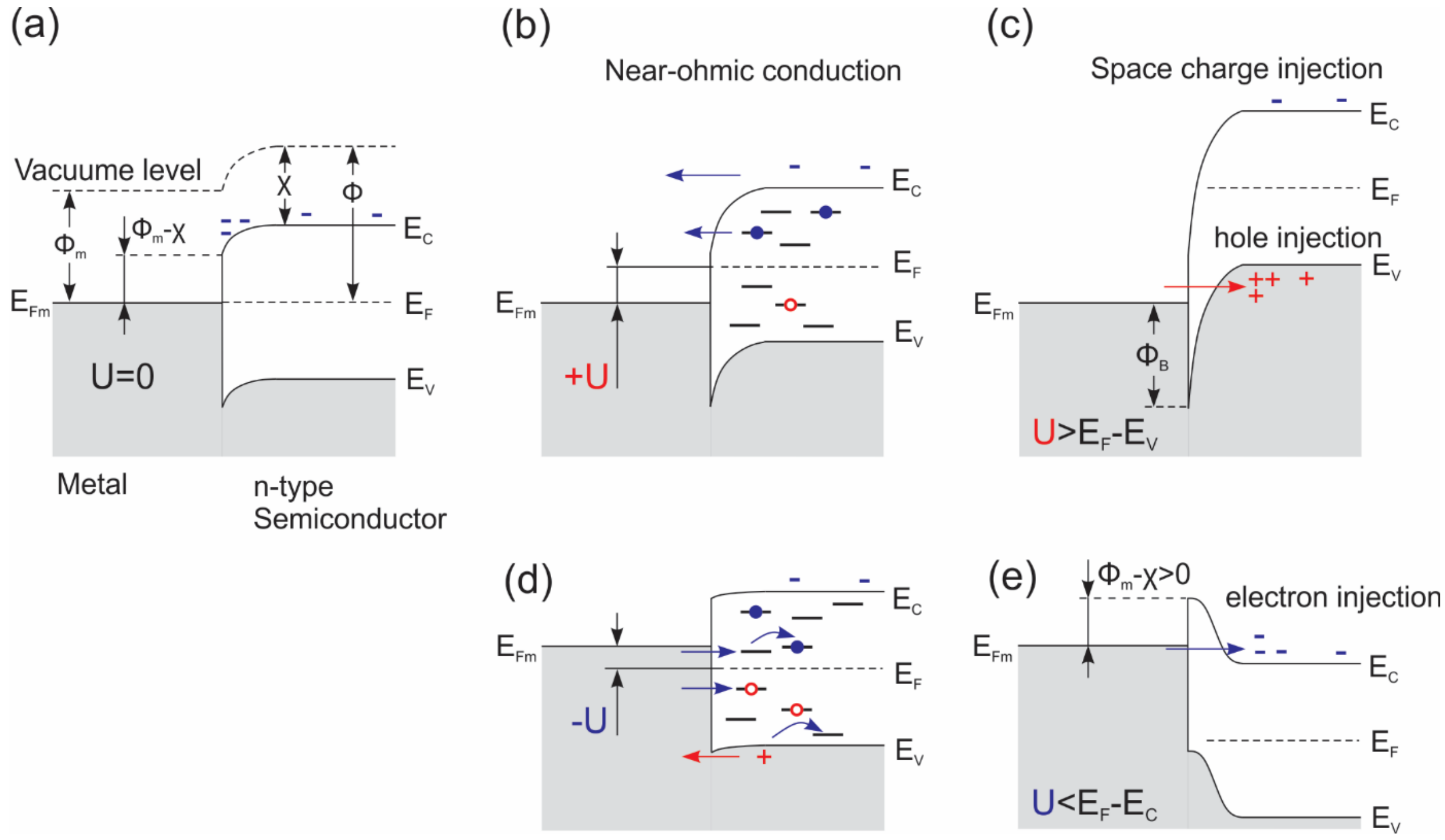

Figure 3. Energy level diagrams for a contact between a metal and n-type semiconductor. (a) metal/semiconductor junction in equilibrium at zero bias voltage. (b) and (d) near-ohmic or Poole–Frenkel (at high electric fields) conduction regime with thermally assisted tunneling between surface traps: (b) at $0 \le U < E_F - E_V$ and (d) $E_F - E_C < U \le 0$ bias voltage on the metal. (c) and (e) Space-charge injection regimes: (c) hole injection at a positive and (e) electron injection at a negative voltage on the metal.

However, it should be noted that if the electric field under the probe is sufficiently high with a high density of charge traps in the LMO surface layer, the conduction occurs due to the Poole–Frenkel mechanism with a non-linear I-V dependence [43]. For a probe with a small apex radius (~30 nm), a high electric field is likely to exist in the vicinity of the probe-sample contact even at bias voltages below 1 V. The Poole–Frenkel relation for conductivity is an exponential function of the electric field *E* and temperature *T*:

$$\sigma = \sigma_0 \exp\left(\frac{\beta_{pF}}{2k_B T}\sqrt{E}\right), \tag{1}$$

where $k_B$ is the Boltzmann constant, $\sigma_0$ is the low-field conductivity, and $\beta_{pF}$ is the Poole–Frenkel constant $\beta_{pF} = \sqrt{\frac{q^3}{\pi\varepsilon\varepsilon_0}}$ [55], where $q$ is the elementary charge, $\varepsilon_0$ is the permittivity of free space, $\varepsilon$ is the relative dielectric constant [56–58]. It is worth further noting that for an exponential function with a small power factor $\le 1$, the near-ohmic conduction is an accurate approximation at low voltages (Fig. 4 (b)).

The low-field conductivity in Eq. (1) $\sigma_0$ is a function of the distance from the surface plane into the LMO sample depth. However, for simplicity, here we consider the LMO as a uniform media with a constant averaged $\sigma_0$ and obtained an approximate relation for the Pool-Frenkel current *I* through the point contact vs. applied voltage *U* in a semispherical approximation (see Supplementary 6, Eq. (S 6.9)):

$$I = A_1 \cdot U(1+\frac{A_2}{2}\sqrt{U}) \cdot \exp\left(A_2\sqrt{\frac{U}{1+\frac{A_2}{2}\sqrt{U}}}\right), \tag{2}$$

where $A_1 = 2\pi a\sigma_0$ and $A_2 = \dfrac{\beta_{pF}}{2k_B T\sqrt{a}}$. This relation is useful because the values of the contact radius

$a = \left(\dfrac{\beta_{pF}}{2k_B T \cdot A_2}\right)^2$ and the low-field conductivity $\sigma_0 = \dfrac{A_1}{2\pi a}$ can be estimated from the parameters $A_1$ and $A_2$ treated as fitting parameters. The results of fitting of the I-V curves in Fig. 4(a) with Eq. (2) in a voltage range from 0 V to 1V are shown in the inset in Fig. 4(a). They yield $a$ from 4 nm to 8 nm and $\sigma_0$ from 3×10$^{-6}$ S/m to 6.4×10$^{-6}$ S/m. The obtained contact radius values agree well with those calculated with use of the Hertzian contact mechanics model (Eq. (S 7.1), Table S7).

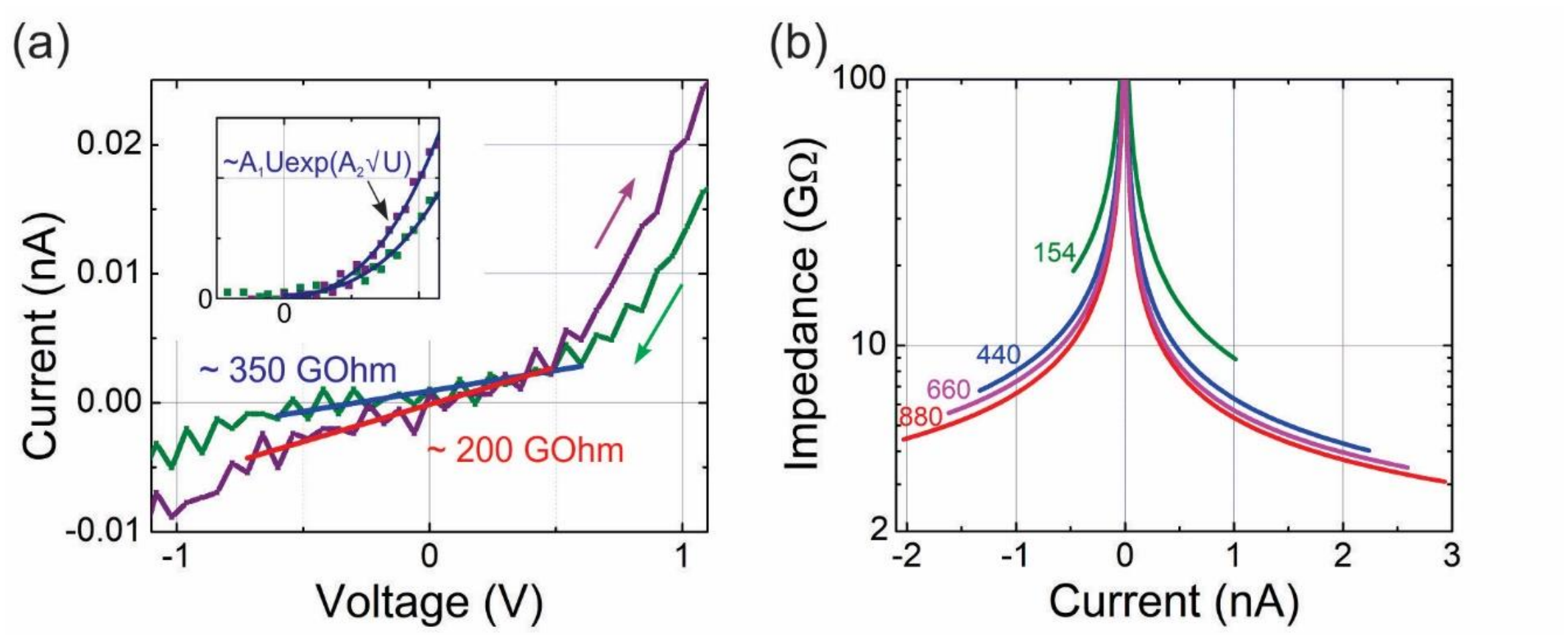

Figure 4. (a) I-V characteristic with near-ohmic conduction in a voltage range from -1 V to +1 V at the back electrode. Inset: fitting of the data in (a) with Eq. (2) in the voltage range from 0 V to 1V. The corresponding fitting parameters and evaluated conductivity are: $A_1$ = (0.22 – 0.07) pS, $A_2$ = 5.7 – 8.4 V$^{-1/2}$, and at $a$ = 6 nm $\sigma_0 = \dfrac{A_1}{2\pi a}$ = 2×10$^{-6}$ S/m to 6×10$^{-6}$ S/m. (b) Probe-sample junction static (chordal) impedance as a function of current for different probe-sample loading forces: 154 nN (green), 440 nN (blue), 660 nN (purple), and 880 nN (red). Measurements in the vicinity of zero voltage give a maximal impedance of ≥ 220-350 GΩ as illustrated by the linear fits in (a).

In turn, the estimated value of $\sigma_0$ is more than one order of magnitude lower than the typical LMO bulk conductivity in the literature (see Table S5). We suppose that this value reflects the presence of a highly resistive surface layer on the LMO sample and corresponds to the surface layer conductivity.

**1.3** ***Conductivity of the surface layer***

The thickness of the high-resistance surface layer, $t$, can be estimated from the value of the impedance in the vicinity of the zero bias voltage. Near zero bias, Eq. (2) simplifies to a linear relation $I \approx A_1 \cdot U$, where $1/A_1$ has the meaning of the impedance limit at zero bias, which is larger than the impedance 220-350 GΩ seen in the data in Fig 4(a). Let us define the conductivity of the surface layer as $\sigma_0'$ and the conductivity in the bulk beneath the surface as $\sigma_0$. Then, the resistance of the tip-sample junction can be found as a resistance of the point contact with a radius $a$. For estimates, we use a semispherical approximation for the contact shape. The used approach is based on substitution of a nearly flat point contact by a hemispherical depression of a radius $a$ on the sample surface with the sphere center in the plane of the sample surface. With $\sigma_0' << \sigma_0$, we get:

$$R = \frac{1}{2\pi}\int_a^\infty \frac{dr}{r^2\sigma(r)} = \frac{1}{2\pi}\left(\int_a^{a+t} \frac{dr}{r^2\sigma_0'} + \int_{a+t}^\infty \frac{dr}{r^2\sigma_0}\right) \approx \frac{1}{2\pi a\sigma_0'}\left(1 - \frac{a}{a+t}\right),$$

where $r$ the is radial distance in the spherical coordinate system with the origin at the contact center in the plane of the sample surface. Equating the last expression to the measured impedance, we can estimate the thickness of the surface layer as $t = \frac{a}{1/(2\pi aR\sigma_0') - 1} > \frac{a}{1/RA_1 - 1}$. Here, we assume that $1/R > 2\pi a\,\sigma_0' > A_1$, which is equivalent to the conductivity dropping from the LMO bulk towards surface within a thin surface layer. With $R$ = 220 – 350 GΩ, and for $A_1$ = 0.07 pS, we obtain $t >$ 0.02$a$.

Examination of the static (chordal) junction impedance, $\frac{U}{I}$, as a function of the probe-sample loading force and current (or voltage) sheds additional light on the conduction mechanism. As seen in Fig. 4(b), the static impedance increases from 3 GOhm to 350 GOhm with decreasing current for different loading forces. Corresponding contact radii, $a$, for different loading forces are ranged from about 3.3 nm to 5.8 nm (see Supplementary 7, Table S7). The data in Fig. 4(b) with a strong dependence of the impedance on the bias voltage at different probe-sample loading forces corroborate the presence of a highly resistive layer at the LMO surface. This layer is responsible for the essential part of the voltage drop in the system and nonlinear behavior of the junction impedance. The high resistivity of the surface layer assumes that a higher trap density is expected close to the sample surface as compared to that in the bulk. To support this, EDXA experiments (Supplementary 1) show that the surface layer of the LMO sample contains Mn and O in a ratio 2/5. This means that the composition of the LMO surface layer slightly differs from that of the bulk and can contain complexes with oxygen excess, as well as multiple structural defects, which can serve as trapping sites.

The high resistivity of the surface layer can be explained by the multiple trap-and-release (MTR) model originally introduced by Shur and Hack [59] for amorphous silicon. According to the model, trap states with energies within a few $k_B T$ near the mobility edge (shallow traps) are characterized by a finite trapping time. After being trapped for a characteristic time $\tau_{tr}$, a trapped polaron can be thermally activated and released to the conduction band. As a result, the effective mobility $\mu_{eff} = \mu_0 \frac{\tau}{\tau + \tau_{tr}}$ in the surface layer is reduced in comparison with its intrinsic bulk value $\mu_0$. Here, $\tau_{tr}$ is the average trapping time on shallow traps, and $\tau$ is the average time that is spend by a polaron diffusively traveling between consecutive trapping events. In the case $\tau_{tr} >> \tau$, the charge transport is dominated by trapping, and $\mu_{eff} = \mu_0 \frac{\tau}{\tau_{tr}} \propto \exp\left(-\frac{W_{tr}}{k_B T}\right)$. Therefore, the surface layer with a higher trap density should have a higher resistivity compared to the bulk, and its decrease at high currents, seen as a decrease in the static impedance of the junction in Fig. 4(a), can be attributed to the SCLC through the sample surface layer.

To summarize the results of this section, we reiterate that in the voltage range from -1 V to 1 V, the dominating mechanisms defining the character of C-AFM I-V curves is the Poole–Frenkel conduction, and at higher voltages, the dominating mechanism is SCLC in the surface layer.

## 2. Interpretation of I-V curves

### 2.1 *Model*

With the goal to establish suitability of the I-V curve mapping to determine the Li-concentration-dependent conductivity of the subsurface region of the LMO sample, we further elaborate the results of the previous section. We employ the SCLC theory developed for insulators and semiconductors in [40–42,60] adapting it for the case of the C-AFM measurements. For detailed analysis of the I-V curves, we divide the sample into surface and bulk regions. It is sufficient to allow the trap number density $N_t$ (the number of traps per unit volume) to depend on the position relatively to the LMO surface, so that $N_t$ in the region close to the junction is much larger than the bulk value. A schematic of the surface charge traps uniformly distributed within the surface layer is shown in Fig. 5. Figure 6 presents a simplified model of the tip-sample junction with a semi-spherical distribution of the electric field $E$. The full impedance of the system is a sum of an impedance of the surface region, $R_s = \tilde{R}_s I^{\frac{1}{\alpha_1}-1}$, which can also include the interface resistance, and an impedance of the bulk region, $R_b = \tilde{R}_b I^{\frac{1}{\alpha_2}-1}$, where $\tilde{R}_s$ and $\tilde{R}_b$ are coefficients becoming equal to the resistance in the ohmic conduction regime at $\alpha_1 = 1$ and $\alpha_2 = 1$. Neglecting the interface resistance between the metallic tip and LMO surface, the applied bias voltage can be expressed as a sum of voltage drops across the surface layer and bulk, $U = U_s + U_b$:

$$U = \tilde{R}_s I^{\frac{1}{\alpha_1}} + \tilde{R}_b I^{\frac{1}{\alpha_2}}. \qquad (3)$$

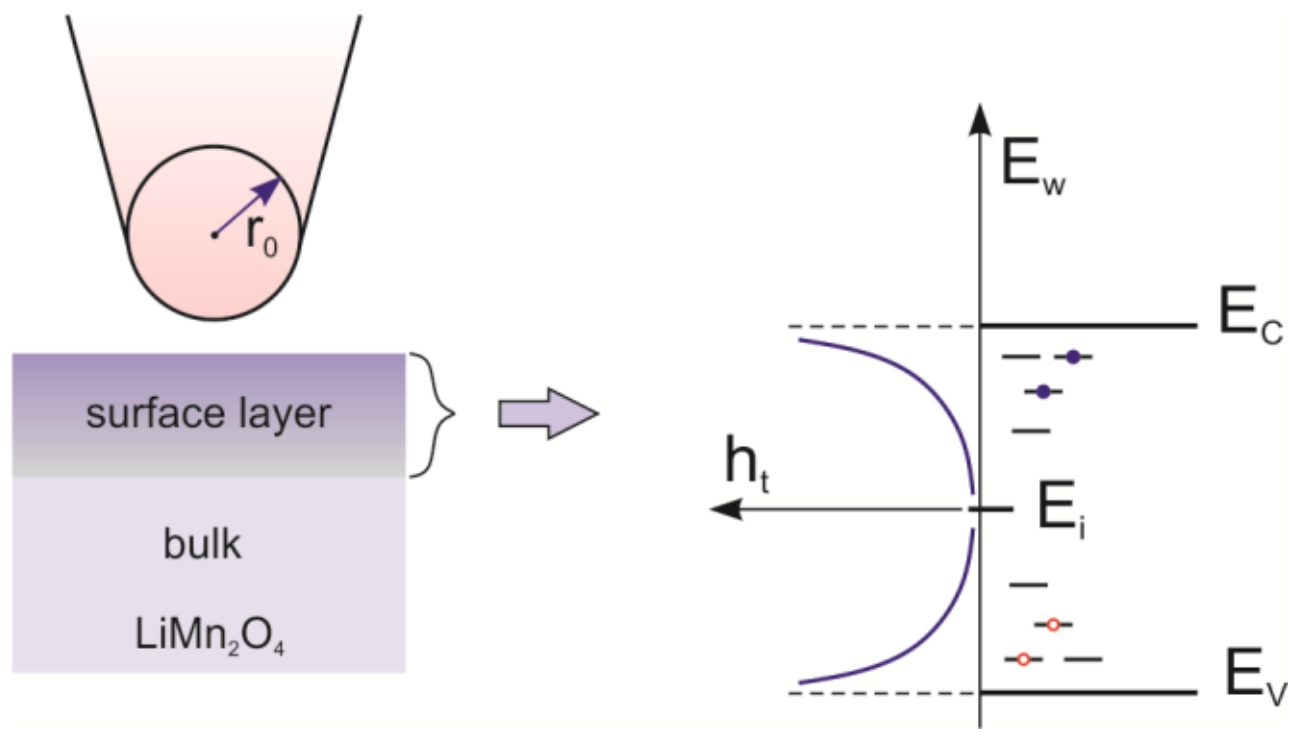

Figure. 5. Left: Schematic of the sample near the surface region with a surface layer filled with charge traps. Right: Schematic of the energy distributions, $h_t(E)$, for donor-like ($E_V \le E_w \le E_i$) and acceptor-like ($E_i \le E_w \le E_C$) traps. The traps are exponentially distributed over energy.

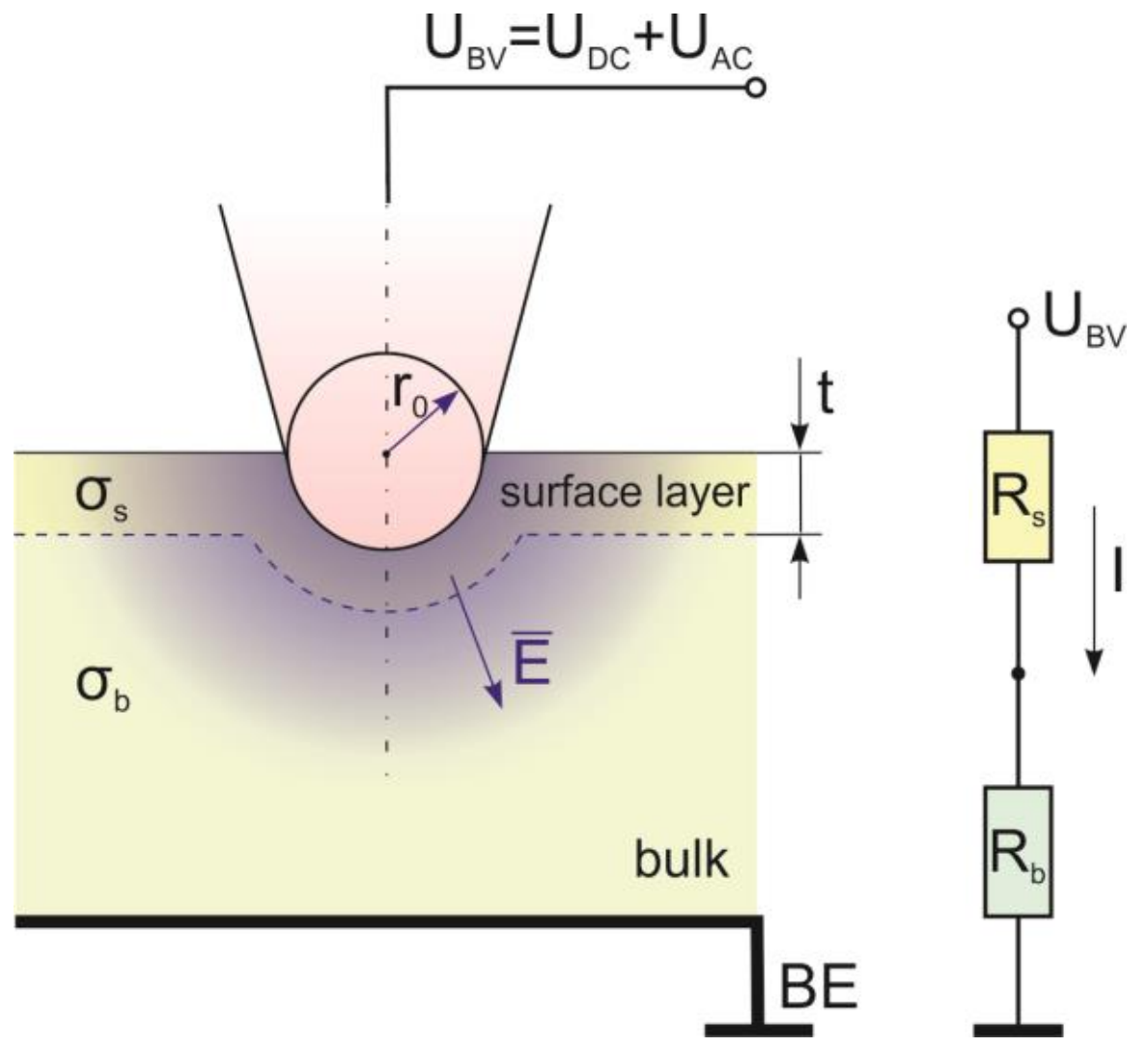

Figure. 6. Schematic of the AFM probe in contact with the LMO particle. A spherically symmetrical approach is used for the calculation of the potential distribution in the half-space near the tip. The equivalent electric circuit (right) includes: $R_s$ is resistance of the surface layer of a thickness *t*, including interface resistance between the metallic tip and LMO surface, $R_b$ is the spreading resistance of the bulk.

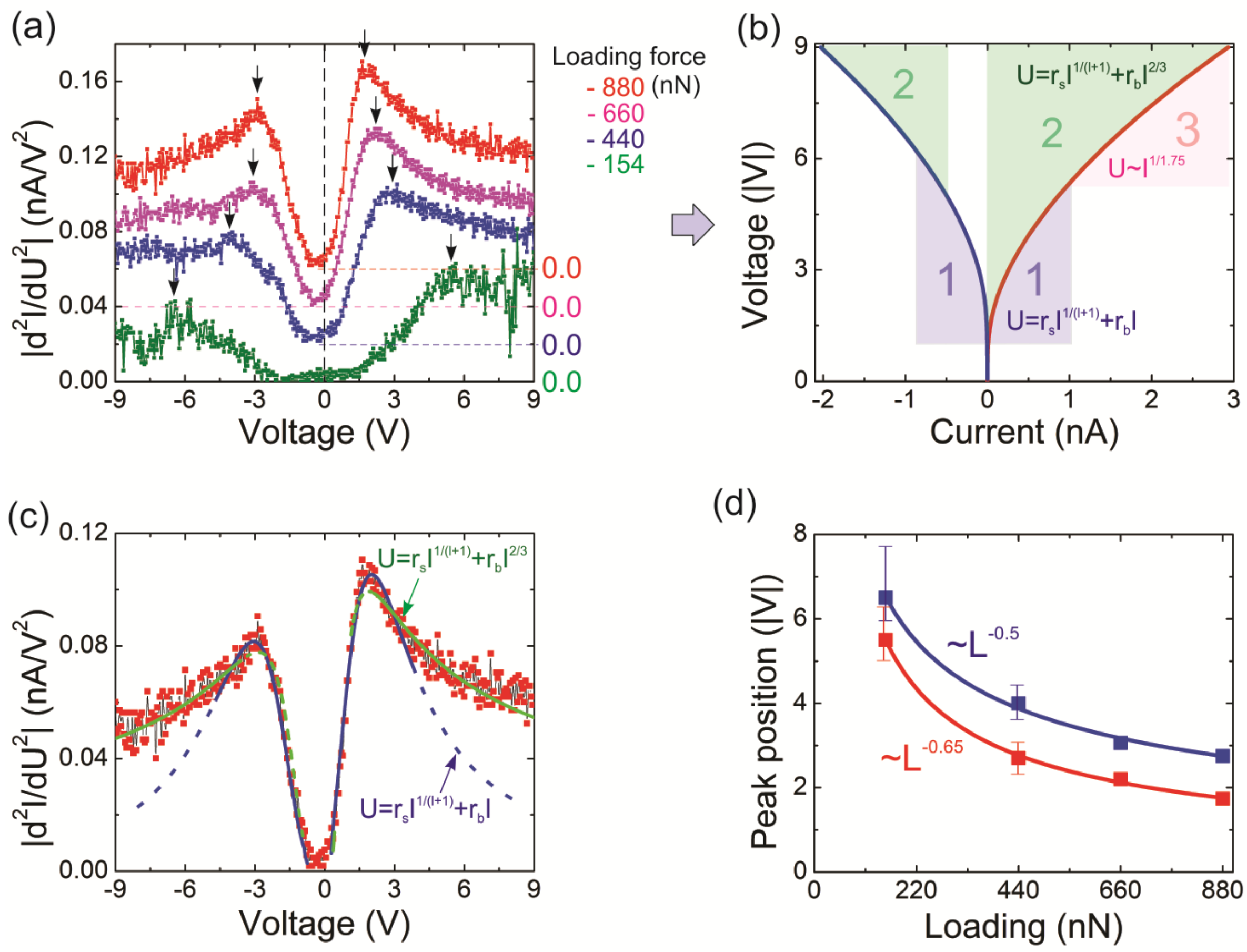

Figure. 7. Second derivative $\frac{\partial I^2(U)}{\partial U^2}$ and its analysis. (a) $\frac{\partial I^2(U)}{\partial U^2}$ as function of the voltage at different probe-sample loading forces (the curves are vertically offset for clarity). Maximum positions (marked by black arrows) shift to higher voltages with decrease of the loading force. (b) Inverse I-V dependence obtained by integration of the second derivative for the 880 nN loading force. Blue line is for negative bias voltages, and red line is for positive voltages. The results of fitting with Eq. (3) are presented in the three color-shaded regions: region 1 with one of the resistive elements in the equivalent circuit in Fig. 6 being linear (it is assumed that for bulk $\alpha_2 \approx 1$), region 2 with fractional values of both the power exponents for the elements with $\alpha_2 \approx 3/2$ for the bulk, and region 3, where the data can be fitted only with a single non-linear component (i.e., $\tilde{R}_s$ or $\tilde{R}_b \approx 0$). (c) fitting of the second derivative data for the 880 nN loading force with the second derivative of Eq. (3) in regions 1 and 2 indicated in (b). Position of the maximum of the second derivative indicated by the black arrows in (a) as a function of the probe-sample loading force. Blue line corresponds to the negative bias voltages, and red line corresponds to positive voltages.

### 2.2 *Results of data fitting*

The measured I-V curves, second derivative $\frac{\partial I^2(U)}{\partial U^2}$ curves, and results of the curve fitting with Eq. (3) are shown in Fig 7. Importantly, a prominent property of the second derivative curves is the maxima observed for both polarities as marked with arrows in Fig. 7(a), which displays the second derivative curves for different probe-sample loading forces. The well-defined maxima facilitate the quantification of the model parameters from the experimental data. As seen in Fig. 7(c), the maxima peaks are well fitted with Eq. (3). The fitting parameters for the curves in Fig. 7(a) in region 2, as indicated in Fig 7 (b), are listed in Table 1.

Table 1. Fitting parameters for I-V curves in Fig. 7(a). The fitting was performed with Eq. (3).

| Loading force, nN | Probe bias voltage polarity (referenced to BE) | $\alpha_1$ | $\alpha_2$ | $\tilde{R}_s$, GΩ·nA$^{(1-1/\alpha1)}$ | $\tilde{R}_b$, GΩ | Current at $\frac{\partial I^2(U)}{\partial U^2}$ maximum, nA | Surface/bulk impedance ratio at $\frac{\partial I^2(U)}{\partial U^2}$ maximum |
|---|---|---|---|---|---|---|---|
| 880 | - | 3.06 ±0.03 | 1 ±0.03 | 4.34 ±0.06 | 0.99 ±0.06 | 0.165 | 14.8 |
| | + | 4.18 ±0.04 | 1 ±0.02 | 5.37 ±0.05 | 1.45 ±0.04 | 0.17 | 14.1 |
| 660 | - | 3.03 ±0.02 | 1 ±0.02 | 4.67 ± 0.04 | 1.02 ± 0.04 | 0.17 | 14.9 |
| | + | 3.92 ±0.02 | 1 ±0.02 | 5.92 ± 0.03 | 1.61 ± 0.03 | 0.165 | 14 |
| 440 | - | 3.15 ±0.02 | 1 ±0.02 | 5.37 ± 0.05 | 0.93 ± 0.05 | 0.26 | 14.5 |
| | + | 3.76 ±0.02 | 1 ±0.02 | 6.9 ± 0.05 | 1.37 ± 0.05 | 0.26 | 14 |
| 154 | - | 3.42 ±0.02 | 1 ±0.02 | 8.2 ± 0.04 | 0.84 ± 0.04 | 0.6 | 14.1 |
| | + | 3.74 ±0.07 | 1 ±0.1 | 10.3 ± 0.2 | 2.11 ± 0.2 | 0.24 | 14 |

The spreading resistance of the bulk $\tilde{R}_b$ is the parameter reflecting the Li-ion concentration under the probe, and it requires accurate evaluation if mapping of ion distribution is desired. The impedance ratios between surface and bulk regions in the last column of Table 1 were calculated as $\frac{\tilde{R}_s I_p^{\frac{1}{\alpha_1}}}{\tilde{R}_b I_p^{\frac{1}{\alpha_2}}}$, where $I_p = I(U_{peak})$ in the current at the bias voltage $U_{\text{peak}}$ corresponding to the second derivative maximum vs. bias voltage.

We note that in the fitting results in Table 1, the power index for the bulk is $\alpha_2 \approx 1$, which points to the ohmic charge transport in the bulk independently of the tip-sample loading force [42]. As shown in Supplementary 8 (Eq. (S 8.2)), Eq. (3) with $\alpha_2 = 1$ allows determination of $\tilde{R}_b$ from measured resistance of the surface layer, $\tilde{R}_s$, and $I_p$ using relatively simple expressions. Namely, the impedance ratio becomes a function of only $\alpha_1$:

$$\frac{I_p^{\frac{1}{\alpha_1}} \tilde{R}_s}{I_p \tilde{R}_b} = \frac{\alpha_1(2\alpha_1 - 1)}{\alpha_1 - 2}, \tag{4}$$

and $\tilde{R}_s$ and $\tilde{R}_b$ are related as (Supplementary 8, Eq. (S 8.1)):

$$\tilde{R}_b = \frac{\alpha_1 - 2}{\alpha_1(2\alpha_1 - 1)} \tilde{R}_s I_p^{\left(\frac{1-\alpha_1}{\alpha_1}\right)} . \tag{5}$$

However, it should be noted for the data of the last column of Table 1 that the impedance of the surface layer is more than one order of magnitude greater that the impedance of the bulk. In view of the determination of the local Li-ion concentration from the measurements of the I-V curves, this result is unfavorable since the relatively large resistance of the surface layer strongly masks variations of the bulk resistance.

Values of $\alpha_1$ in Table 1 are in a range from about 3 to 4, which demands a more detailed consideration. These values unambiguously indicate the SCLC charge transport in the surface layer. The space charge is localized within the high-impedance surface layer. To show that we note that the size of the space charge region in a uniform medium under the probe, $x_c$, can be estimated as a distance traveled by the injected charge within the Maxwell relaxation time, $\tau_M = \frac{\varepsilon\varepsilon_0}{\sigma}$, i.e., the time required for the injected charge to dissipate into the surrounding conducting medium under action of its own electric field, where $\varepsilon_0$ is vacuum permittivity, $\sigma$ is conductivity, and $\varepsilon$ is the medium permittivity [61–67]: $x_c = \mu E \tau_M$, where $\mu$ is the charge carrier mobility. For a given probe current *I*, in the semispherical approximation, the electric field as a function of the distance $r$ from the center of the probe apex sphere (Fig. 6) can be determined from Ohm's law:

$$E_r(r) = \frac{I}{2\pi r^2 \sigma_{e^-}(r)} = \frac{I}{2\pi r^2 q \mu_{e^-} \cdot n(r)}, \tag{6}$$

where, $q$ is the elementary charge, $\sigma_{e^-}(r)$ is the LMO electron conductivity that is further expressed through the electron mobility, $\mu_{e^-}$, and electron concentration, $n(r)$. Using the parameter values for LMO given in Table S5, and assuming $n_{e^-} \approx n_{Li}(r)$, the size $x_c$ can be estimated as $x_c \approx 20$ nm in LMO bulk with $I = 1$nA and $r = 6$ nm corresponding to the contact radius. Since the impedance of the surface layer is much larger than that of the bulk according to the fitting results in the last column of Table1, $\tau_M$ and $x_c$ for the surface layer are much larger than those for LMO bulk, and, therefore, the injected charge fully fills the thickness of the surface layer. Based on the estimate of the thickness of the surface layer $t > 0.02a$ obtained in the previous section, we may take $t \approx 0.1a$, in the model in Fig.

6. Since the tip-sample contact radius $a$ ~ 6 nm is significantly larger than $t$, the one-dimensional model developed in Refs. [34,54] can be used for the description of SCLC in the surface layer. In the case when the space-charge region is fully contained within the thickness of the surface layer and charge traps are exponentially distributed over energy, the one-dimensional model yields a power law dependence of the current on voltage, $I \sim V^{l+1}$ with $l \geq 1$ [40,42]. Hence, we associate the values of $\alpha_1$ in Table 1 with the SCLC regime in the planar surface layer: $\alpha_1 = l+1$, where $l > 1$.

### 2.3 *High current*

At larger current and applied bias values, the relative contribution of the surface layer in the resistance drops, and the charge transport in the bulk becomes more governed by the SCLC mechanism. Namely, when concentration of the injected charge achieves a value equal to or higher than the intrinsic thermal equilibrium charge carrier concentration in the bulk, the space charge extends into the bulk where a spherically symmetric three-dimensional model can be used for the description of the space charge transfer (Supplementary 9). For the trap-free case [42] or traps exponentially distributed in energy (Supplementary 9), in the spherically symmetric case, the current will follow the 3/2 power law, $I \sim U^{\frac{3}{2}}$, and the power index for the bulk in Eq. (3) becomes $\alpha_2$ = 3/2. A consideration of transition between ohmic and SCLC regimes in the bulk can be performed using the regional approximation method [42]. The size of the space charge region for the one-dimensional case is a function of conductivity, $x_c \sim \frac{J}{\sigma}$ (Supplementary 9, Eq. (S 9.25)):

$$x_c = \frac{J\varepsilon\varepsilon_0}{A\sigma n^l}\frac{l}{l+1}, \tag{7}$$

where $J$ is the current density, $l \geq 1$, and $A$ is the ratio of the number of states per unit volume in the conduction band (for electrons) or valence band (for holes) to the concentration of charge traps (see Supplementary 9). Consequently, $x_c$ is two orders of magnitude larger for the highly resistive surface layer than for the bulk, and the space charge can be confined within the surface layer approximately two decades of current values, i.e., from ~0.01 nA to ~1 nA (Fig. 7 (b), region 1).

### 2.4 *Second derivative maximum vs. probe-sample loading force*

Now we consider the interpretation of the shift of the second-derivative maximum as a function of the probe-sample loading force $L$ (Fig. 7(d)). This parameter is important since apparently, this shift is a result of redistribution of the full resistance of the probe-sample system between the surface and bulk components due to the loading-force-dependent radius of the probe-sample contact. The position of the maximum of the bias voltage, $U_p$, (Fig. 7(a)) can be expressed from Eq. (3) with $\alpha_2 = 1$ and Eq. (4) (Supplementary 8, Eq. (S 8.4)) as:

$$U_p = \left( \frac{(\alpha_1 - 2)}{\alpha_1(2\alpha_1 - 1)} \frac{\tilde{R}_s^{\alpha_1}}{\tilde{R}_b} \right)^{\frac{1}{\alpha_1 - 1}} \frac{2\alpha_1^{\ 2} - 2}{\alpha_1(2\alpha_1 - 1)} . \quad (8)$$

Assuming that the surface layer thickness $t << a$, where $a$ is the contact radius, we can estimate the contributions to the full impedance from the bulk and surface layers as functions of the contact radius as $\tilde{R}_b \sim \frac{1}{a}$ [68] and $\tilde{R}_s^{\alpha_1} \sim \frac{1}{a^2}$, respectively. Then, Eq. (8) can be simplified to $U_p \sim a^{-\frac{1}{\alpha_1 - 1}}$. Using Eq. (S 7.1) for the Hertzian contact radius at a loading force $L$, we obtain:

$$U_p \sim L^{-\frac{1}{3(\alpha_1 - 1)}} . \quad (9)$$

With Eq. (8), we find that at a positive voltage on the back electrode and $\alpha_1 \approx 3$, $U_p \sim L^{-0.17}$; in turn, at a negative voltage on the back electrode and $\alpha_1 \approx 4$, $U_p \sim L^{-0.11}$. The experimental values of the power exponents for the loading forces for the plots in Fig. 7(d) are higher than estimated with Eq. (9). This can be due to a deviation of the contact radius from that predicted by the Hertzian model. The assumption $\tilde{R}_b \sim \frac{1}{a}$ is an approximation inaccurately accounting for the thickness of the surface layer, which can also contribute to the discrepancy.

It should be noted that the observed dependence of the maximum position on the loading force in (Fig. 7 (a), and (d)) cannot be attributed to the trap energy distribution in the surface layer. Suppose that for a fixed trap energy distribution in the surface layer, the position of the maximum is determined solely by the potential drop across the surface layer $U_0$ and, therefore, can be found from an equation $\tilde{R}_S I^{\frac{1}{\alpha_1}} = U_0$, where $U_0$ is determined by the trap energy distribution only and does not depend on the current. Expressing the current as $I = \left( \frac{U_0}{\tilde{R}_S} \right)^{\alpha_1}$ and substituting it into Eq. (3), we obtain: $U_p = U_0 \left( 1 + U_0^{\alpha_1 - 1} \frac{\tilde{R}_b}{\tilde{R}_S^{\alpha_1}} \right)$. Estimating $\tilde{R}_b \sim \frac{1}{a}$ and $\tilde{R}_s^{\alpha_1} \sim \frac{1}{a^2}$ and using the Hertzian relation for the contact radius, we get: $U_p \sim L^{\frac{1}{3}}$. The last relation predicts the opposite direction of the maximum position shift in response to a change of the loading force.

### 2.5 *Discussion of the model results*

The presence of highly resistive surface layer changes the dynamics of ion redistribution and significantly reduces the apparent electromechanical SPM signal associated with the ionic motion in the sample (ESM signal) [7]. It should be noted that the impedance ratios at the position of the second derivative peaks of the I-V curves (Fig. 7, Table 2) suggest that the surface layer in the SCLC regime takes a 14 times lager voltage drop than the bulk in the samples used in this work. This result was obtained within the model used for region 1. With increase of the bias voltage, the fraction of the voltage drop across the surface layer decreases. For the model used for higher currents in region 2 (Fig. 7(b)), the impedance ratios at the boundary between regions 1 and 2 (Fig. 7(c)) have ~2-3 times lower values. The mismatched values for the voltage drop across the surface region under different charge transfer regimes (ohmic and SCLC) in the bulk indicate a deficiency of the simplified model used in the analysis. Apparently, the simple separation of the sample into two spatial domains (Fig. 6) does not properly describe the transition between the different regimes. However, the model captures well the behavior of I-V curves at different voltages and loading forces. More accurate analysis requires not only consideration of the spatial trap distribution but also account of anisotropy of the conduction in the surface layer, which can be higher in the lateral direction in comparison with the normal to the surface because charges injected from the metal conduction band (Fig. 3(d)) can spread along the surface layer by thermally assisted tunneling between surface traps. Charge spreading through a surface conduction is known for semiconductor materials, for example for a Si(111)-7x7 surface [69]. The existence of surface states also assumes an additional resistive barrier between the surface and bulk. In the analysis performed above, we have considered only a specific case with a constant electron hopping mobility. More accurate calculations would require a consideration of electron mobility as a function of strain and Li concentration.

### 2.6 *A note: Effect of charge injection on the material strain*

The SCLC injection in the surface layer can result in mechanical strain induced by electron and hole concentration changes similar to the Vegard effect for concentration of ions [70,71]. Normally, this effect is about one order of magnitude lower than the Vegard strain induced by ion redistribution. However, due to high density of the injected charges, the electron and hole contributions can be comparable with the ionic contribution in sample response measured, e.g., in the ESM signal. For example, the ESM response in ceria observed at room temperature, when the ionic transport is sufficiently suppressed, in Ref. [72] can be attributed to this effect. Dynamics of the injected holes and electrons is defined by the Maxwell relaxation time, which is several orders of magnitude shorter than the ionic diffusion times. The charge-injection-induced strain can also explain the existence of the background offset observed in the low-frequency ESM signal in Ref. [25] (Fig. 9), that was higher the noise level measured in amorphous silicon (Fig. 9, green circles) [25]. The charge-injection-induced strain can reasonably explain the appearance of this background offset in the ESM signal.

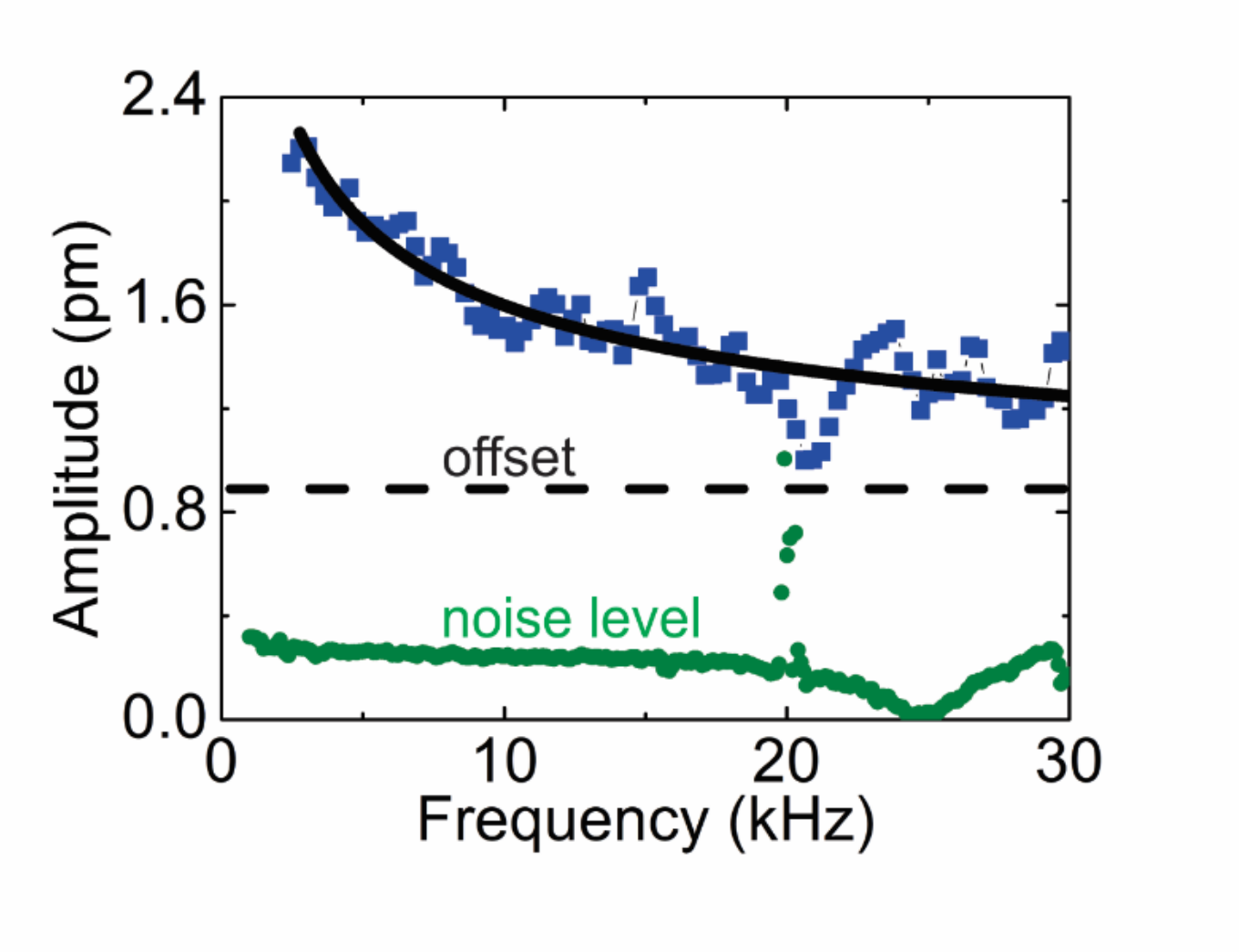

Figure 9. A representative frequency-dependent ESM signal measured on a LMO cathode (blue squares) in Ref. [25]. The black curve is a fit. The dashed line indicates the level of the background signal. The green balls indicate the noise level measured in amorphous silicon. Adapted from Ref. [25].

### 3. Effect of the Li-ion redistribution on the I-V curves

In the previous section, we established that in the case of the ion blocking (i.e., transparent for electrons and holes only) probe-sample interface, the fitting coefficients in Eq. 5, $\tilde{R}_s$ and $\tilde{R}_b$, describe the SCLC regime in the surface layer and the bulk conductivity outside of the surface layer, respectively. In this section, we focus on the bulk conductivity outside of the surface layer for the low-voltage/low-current range marked in Fig. 7(b) as region 1. Region 1 is characterized by the presence of the peak in second derivative $\frac{\partial I^2(U)}{\partial U^2}$, which, we assume, is a combined result of SCLC conductivity in the surface layer and ohmic conductivity in the bulk. The electronic current distribution under the condition of the ohmic conduction determines the electric field causing the Li-ion migration in the bulk. Due to the ohmic conduction, we can apply the charge neutrality condition in the bulk and use the linear drift-diffusion equations to evaluate the ionic concentration under the probe. Here we assume that the spatial domains with SCLC (surface layer) and ohmic (bulk) conduction are separated by a sharp boundary, which is an approximation of the model. We consider the effects of the Li-ion redistribution in the electric field of the ion blocking probe on the I-V curves of the probe-sample contact, provide expressions for estimation of the point contact resistance with account of the Li-ion stationary concentration profiles in the non-uniform electric field in the bulk, and make an estimations of the amplitude of the Li-ion concentration change in response to the probe voltage bias in our experiments.

### 3.1 *I-V curves with Li-ion redistribution in the electric field of the ion-blocking probe*

In the geometry of Fig. 6, distributions of electron current and ion concentration in a uniform media for quasi-stationary conditions can be solved using Ohm's law: $J = \sigma \cdot E_r(r)$, where *J* is the current density and $E_r(r)$ is the radial component of the electric field (in the spherical reference system). We assume the charge neutrality and an approximation $n_{e^-} \approx n_{Li} = n$ [25,30]. Then the conductivity is $\sigma = q\mu_{e^-} \cdot n$, where $\mu_{e^-} \propto \exp\left(-\frac{W(n_{Li})}{k_B T}\right)$ is electron hopping mobility as a function of activation energy $W(n_{Li})$, which is, in turn, a function of Li-ion concentration. According to Refs. [34–36] the activation energy can be increased several times with an increase of the Li-ion concentration and lattice parameter. However, the excited volume under the tip apex is deformation-limited and expands only partly in the upward direction. Due to that, the dependence of electron hopping mobility on Li concentration in the volume under the tip is expected to be significantly weaker than in uniformly lithiated cathodes. Using above arguments, in the following estimations we make a simplification and use a constant value for $\mu_{e^-}$. Then, for the half-space with the total current $I = 2\pi r^2 J$, the resulting electric field as a function of radius is given by Eq. (6). Equation (6) is valid for any spherically symmetric concentration distribution $n(r)$. The balance between diffusion and migration fluxes of the Li ions in the steady state can be written as:

$$D\frac{\partial n}{\partial r} = \pm \mu_{Li} n \cdot E(r), \tag{10}$$

where $\mu_{Li}$ is the Li-ion mobility. The ± sign is determined by the polarity of the electric field. Substituting Eq. (9) into Eq. (10) results in an equation:

$$D\frac{\partial n}{\partial r} = \pm \frac{\mu_{Li} I}{\mu_{e^-} 2\pi q r^2}, \tag{11}$$

that has a solution:

$$n(r) = n_0 \pm \frac{I}{\mu_{e^-} k_B T 2\pi r}, \tag{12}$$

Where we used the Einstein relation $D = \frac{\mu_{Li} k_B T}{q}$.

By stoichiometry, Li-ion concentration is limited from top by a maximal value $n_{\max}$ and from the bottom by a minimal value $n_{\min}$. Radial coordinates corresponding to $n_{\max}$ and $n_{\min}$ at negative and positive polarity of the tip bias voltage are defined using Eq. (12) as $r_-(I) = \frac{I}{2\pi \mu_{e^-} k_B T (n_{\max} - n_0)}$ and $r_+(I) = \frac{I}{2\pi \mu_{e^-} k_B T (n_0 - n_{\min})}$, respectively. In the model, we assume abrupt transitions from the concentration profiles given by Eq. (12) to the constants $n_{\max}$ and $n_{\min}$. The resulting concentration profiles at negative and positive tip bias polarities are schematically shown in Fig. 8.

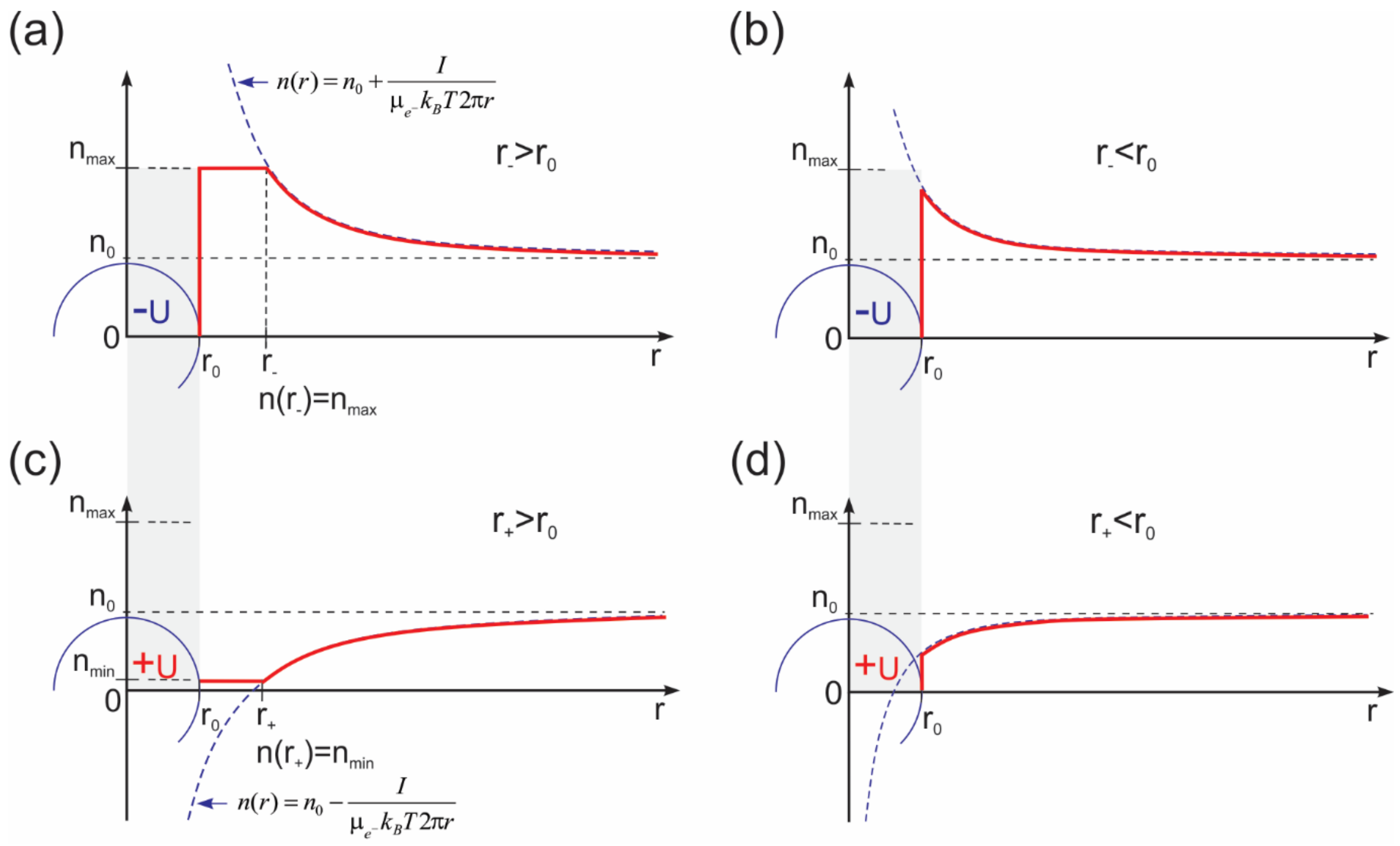

Figure 8. Schematic profiles (red lines) of Li-ion concentration distributions under the tip of a radius $r_0$ in a steady state: (a), (c) for negative voltages on the tip; (b), (d) for positive voltages on the tip. (a), (c) distributions at $r_0 < r_+$ and $r_0 < r_-$, accounting for the steric effect limiting the maximal concentration to $n_{\max}$, and for the minimal concentration $n_{\min} \geq 0$, respectively; (b), (d) distributions at $r_0 > r_+$ and $r_0 > r_-$, respectively.

As seen the concentration profiled of Li ions can be divided into two groups: (i) high-current (Fig. 8 (a), (c)), so that $r_0 < r_+$ and $r_0 < r_-$, i.e., the maximum and minimum concentrations are reached in the sample bulk, and (ii) low-current (Fig. 8(b), (d)) when $r_0 > r_+$, $r_0 > r_-$, i. e., the ion concentration remains between $n_{\min}$ and $n_{\max}$ everywhere in the sample. The contribution from the bulk into the point contact resistances corresponding to these cases for positive and for negative polarities on the tip are calculated in Supplementary 10. The resulting functions $U(I)$ including contributions from the surface layer based on the model in Fig. 6 (Eq. (5)) for both the polarities are:

$$U_+(I) = I^{\frac{1}{\alpha_+}} R_s + IR_{\max} - \frac{k_B T}{q}\left(\frac{n_0}{n_{\min}} - \ln\left(\frac{n_0}{n_{\min}}\right) - 1\right), \quad (13)$$

$$U_-(I) = I^{\frac{1}{\alpha_-}} R_s + IR_{\min} + \frac{k_B T}{q}\left(\frac{n_0}{n_{\max}} - \ln\left(\frac{n_0}{n_{\max}}\right) - 1\right), \quad (14)$$

with

$$R_{\max} = \frac{1}{2\pi r_0 q \mu_{e^-} n_{\min}}, \quad (15)$$

and

$$R_{\min} = \frac{1}{2\pi r_0 q \mu_{e^-} n_{\max}} \quad (16)$$

in the high-current case with $r_0 < r_+$ and $r_0 < r_-$, and:

$$U_+(I) = I^{\frac{1}{\alpha_+}} R_s - \frac{k_B T}{q}\ln\left(1 - \frac{r_+}{r_0}\right), \quad (17)$$

$$U_-(I) = I^{\frac{1}{\alpha_-}} R_s + \frac{k_B T}{q}\ln\left(1 + \frac{r_+}{r_0}\right), \quad (18)$$

in the low-current case with $r_0 > r_+$, $r_0 > r_-$. Expanding of the logarithms in Eqs. (17) and (18) into Taylor series in respect to $r_+/r_0$ <<1 gives a linear term $\frac{k_B T}{q}\frac{r_+}{r_0} = \frac{I}{2\pi r_0 q \mu_{e^-}(n_0 - n_{\min})}$ identical for both the voltage polarities, which indicates symmetry of the bulk conductivity in respect to the voltage polarity at small currents. In turn, the estimations for currents ≥ 0.1 nA yield values for $r_+$ and $r_-$ > 50 nm, that is, much higher than the contact radius, and Eqs. (13) and (14) should be applied.

The Eqs. (13) and (14) differ from Eq. (3) used for fitting the data in Fig. 7 by an additional, third, term depending only on the lithium ion concentration. This term, which is independent of the current, is a non-obvious result of the model and accounts for the ions distribution in the MIEC with a non-zero current. It results in an additional potential drop as a function of the temperature and ion concentration only. With $q_{Li} = |e|$ at room temperature, the factor in the last term in Eqs. (13) and (14) is $\frac{k_B T}{q} = 0.026$ V. Then, the additional voltage offset in Eq. (14) at $\frac{n_0}{n_{\max}} = 0.61$ as about +0.0027 V, while the additional voltage drop in Eq. (13) depends on the ratio $\frac{n_0}{n_{\min}}$ ($10 \geq \frac{n_0}{n_{\min}} > 1$) and can amount to about –0.23 V at $T$=300 K and $\frac{n_0}{n_{\min}} = 10$. These voltage offsets were taken into account in the fitting for parameters present in Table 1, however, the fitting results changed insignificantly.

### 3.2 *Amplitude of the Li-ion concentration change*

Now, we turn to estimation of the variations of the Li-ion concentration in response to the voltage bias on the probe. Equations (13)-(18) can be adapted to the model in Fig. 6 by replacing $r_0$ with $r_{eff} \approx r_0 + t$, where $t$ is the surface layer thickness. As suggested by comparison of Eq. (3) and Eqs. (13) and (14), the values of $\tilde{R}_b$ in Table 1 for negative and positive voltages on the tip can be taken equal to $R_{\min}$ and $R_{\max}$, respectively, and can be used for estimation of maximal and minimal Li-ion concentrations, $n_{\max}$ and $n_{\min}$. However due to complex mechanism of charge transfer at the interface, including the SCLC at the surface layers and ohmic conduction in the bulk, it cannot be done directly with values of $\tilde{R}_b$ in Table 1. Instead, we use ratios of $\tilde{R}_b$'s obtained for positive and negative biases. The effective contact radius $r_0$ for determination of $R_{\min}$ and $R_{\max}$ (defined in a uniform media, Eqs. (15) and (16)) should be associated with the surface layer thickness $t$, that is with the virtual boundary between spatial domains with SCLC conduction and ohmic conduction. However, the simplified model in Fig. 6, dividing the system into two regions, does not consider effects of the real spatial distribution of charge traps $N_t(z)$. For traps uniformly distributed in space, the contact radius $r_0$ can be estimated as $r_0 \sim x_c \sim \frac{1}{n^{l+1}}$, (Eq. (S 9.27)), and, hence, $r_0^+ < r_0^-$ since $n^- < n^+$. It can be assumed that with a realistic trap distribution in the sample, the same relation is valid: $r_0^+ < r_0^-$. As before, superscripts "+" and "-" here refer respectively to the positive and negative bias polarity on the probe. With the account of the difference between $r^+$ and $r^-$, the ratio of the bulk resistivities is:

$$\frac{\tilde{R}_b^-}{\tilde{R}_b^+} \approx \frac{R_{\min}^-}{R_{\max}^+} = \frac{r_0^+ n_{\max}}{r_0^- n_{\min}}. \quad (19)$$

From Table 1, $\frac{\tilde{R}_b^-}{\tilde{R}_b^+} \approx 1.5 - 2$, which can be used as a lower limit for the concentration ratio, i. e., $\frac{n_{\max}}{n_{\min}} > 1.5 - 2$, a very reasonable value. Corresponding estimation for the bulk conductivity at a contact radius of 5.8 nm (see Table 1) is: $\sigma_b \le \frac{1}{2\pi a \tilde{R}_b} = (1.4 - 2.8) \cdot 10^{-2}$ S/m, which is in a good agreement with literature data for the LMO conductivity [61–65].

## Conclusions

It can be concluded that the local I-V curves obtained with C-AFM on $Li_{1-x}Mn_2O_4$ ceramic particles reveal complex charge carrier dynamic processes under the probe, including SCLC charge transport across a highly resistive surface layer and ion redistribution in the bulk. The analysis of the I-V curves indicates that at probe bias voltages below 1 V, Pool-Frenkel conduction dominates in the surface layer, while at higher voltages, the dominating mechanism is SCLC, which explains the power-law character of the I-V curves. In the SCLC regime, the surface layer takes an up to 14 times larger voltage drop than the bulk in the studied samples. The analysis allowed us to estimate ratios of maximal and minimal conductions associated with the ion redistribution in the sample bulk under the probe at different polarities of the probe bias voltage. The presence of the surface layer alters the dynamics of the ion redistribution and significantly reduces the effective electromechanical SPM signal associated with the ionic motion in the sample. The SCLC mechanism of the charge transport in the highly resistive surface layer is reflected in the asymmetry of the experimental I-V curves.

The results and analytical framework of this work can be applied in future for studies of local conductivity as a function of the lithium concentration at surfaces of Li-ion conductors as well as for high-resolution mapping of Li-ion distributions based on measurements of local I-V curves. However, an accurate quantitative characterization of the Li-ion distribution will require further elaboration of the model accounting for a gradual transition of the sample properties from the surface into the bulk and an anisotropy of the surface conduction. Possible applications of the analytical framework developed here are not limited to the mixed ion-electronic conductors and can be applied to a broad range of electrical conductors. It can pave the way to deeper understanding of details of the electronic transport across the probe/sample interface, as well as of effects associated with surface layers in the charge transport, mechanical and other responses measured in SPM.

**Acknowledgments**

This work was developed within the scope of the project CICECO-Aveiro Institute of Materials, UIDB/50011/2020 and UIDP/50011/2020, financed by national funds through the Portuguese Foundation for Science and Technology/MCTES. The work was financially supported by the Portuguese Foundation for Science and Technology (FCT) within the project PTDC/CTM-ENE/6341/2014. It is also funded by national funds (OE), through FCT – Fundação para a Ciência e a Tecnologia, I.P., in the scope of the framework contract foreseen in the numbers 4, 5 and 6 of the article 23, of the Decree-Law 57/2016, of August 29, changed by Law 57/2017, of July 19. The authors thank Daniele Rosato (Robert Bosch, GmbH) for providing samples of Li-battery cathodes and useful discussions. Equipment of the Ural Center for Shared Use “Modern nanotechnology” of the Ural Federal University was used in the experiments.

# Local electronic transport across probe/ionic conductor interface in scanning probe microscopy

K.N. Romanuk[1,2], D.O. Alikin[1], B.N. Slautin[1], A. Tselev[2], V.Ya. Shur[1] and A.L. Kholkin[2]

[1] School of Natural Sciences and Mathematics, Ural Federal University, Ekaterinburg, Russia
[2] Department of Physics and CICECO – Aveiro Institute of Materials, University of Aveiro, 3810-193 Aveiro, Portugal

## Supplementary 1.

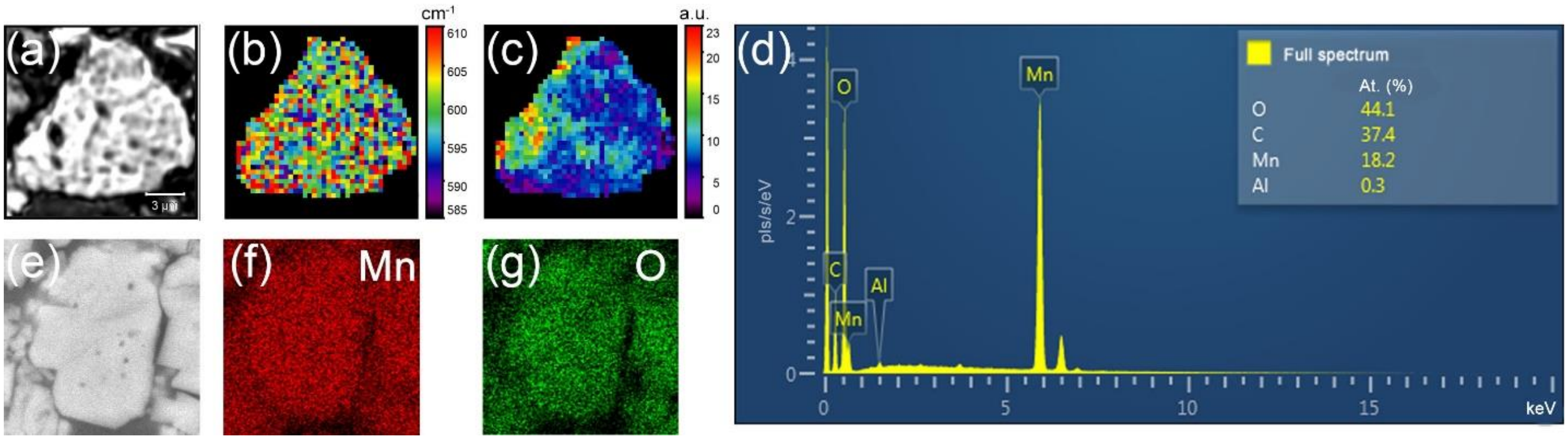

Figure S 1.1. Phase analysis of $LiMn_2O_4$ particle by confocal Raman microscopy and scanning electron microscopy: (a) confocal optical microscopy image of an LMO particle, (b) map of the A1 Raman peak shift, corresponding to the local lithiation state in the particle in (a), (c) map of the $A1_g$ Raman peak intensity reflecting aggregation of the $Mn_3O_4$ phase in the particle in (a), (d) spectra of energy dispersive X-ray analysis (EDXA) with corresponding quantitative elemental analysis, (e) scanning electron image of the area, where the EDXA data in (d) were acquired, (f) map Mn distribution in (e), (g) map of O distribution in (e). The scale bar is equal for all images. The particle in (a) was studied with C-AFM as described in the main text.

According to the EDXA, the surface layer contains Mn and O in a ratio 2/5. It means that the composition of the LMO surface layer slightly differs from bulk and can contain complexes with an excess of oxygen: $Li_5Mn_7O_{16}$, $Li_2MnO_3$, $Li_4Mn_5O_{12}$.

Supplementary 2.

## Heat calculations in the semispherical geometry in a uniform media

The conductivity in $Li_{1-x}Mn_2O_4$ is an activation process depending on the temperature. To establish acceptable current and voltage ranges for I-V and/or SPM measurements, here we perform estimates of heat generation under the probe during charge transfer processes in the semispherical geometry Fig S2.1. In the calculations, we neglected the heat dissipation through the metallic probe (it can be easily accounted for the specified shape of the tip). Due to a significantly higher electric conductivity of the metallic probe in comparison with LMO, we also neglected the Joule heat generation in the probe.

Heat generation and dissipation for a uniform media in the case of the ohmic conduction can be solved analytically.

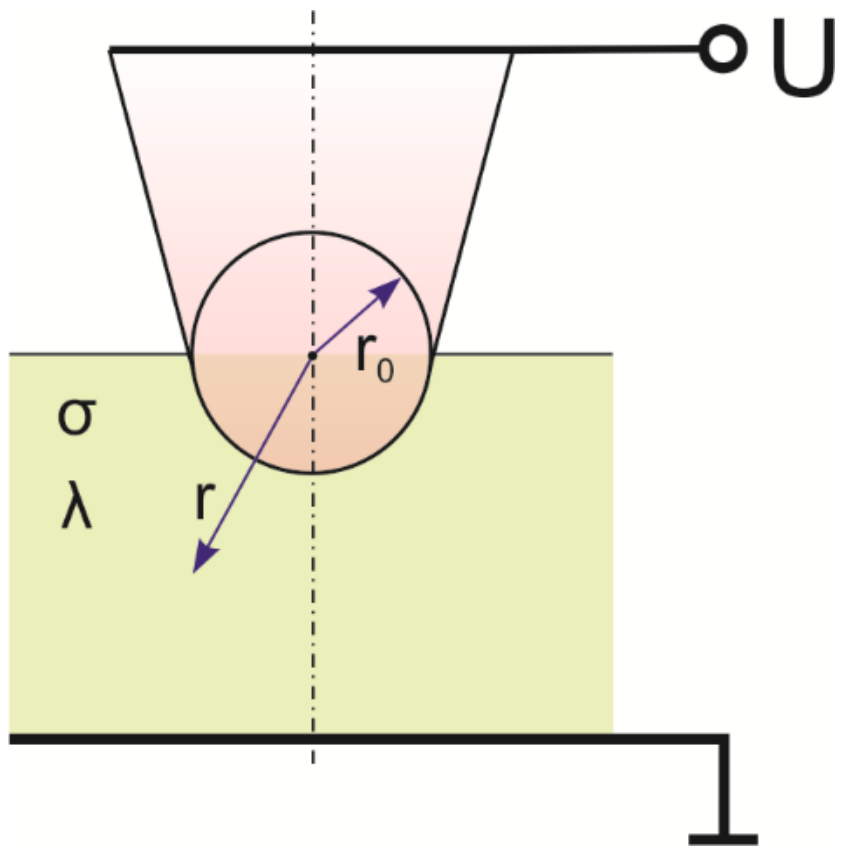

Figure. S 2.1. Schematic of the semispherical geometry for the heat calculations in a uniform media with electric and heat conductivities $\sigma$ and $\lambda$ .

The Joule heat generation in the differential form as a function of the current density is:

$$\omega(r) = \frac{j_r^2(r)}{\sigma(r)} . \qquad \text{(S 2.1)}$$

Using the continuity condition, we can express the current density in the semispherical geometry as:

$$j_r(r) = E_r(r)\sigma_{e^-}(r) = \frac{I}{2\pi r^2} , \qquad \text{(S 2.2)}$$

which is valid for any spherically symmetric distribution of $n_{e^-}(r,t)$. Then, according to Eq. (S 2.1) the Joule heat generation as a function of radius $r$ is:

$$\omega(r)=\frac{I^2}{4\pi^2 r^4 \sigma_{e^-}(r)}. \quad \text{(S 2.3)}$$

The specific heat capacity $C_P$, thermal diffusivity $\alpha$, thermal conductivity $\lambda$, density $\rho$, and electrical conductivity $\sigma$ for $LiMn_2O_4$ are listed in Table S.2.1:

Table. S.2.1.

| Property of $LiMn_2O_4$ | Symbol | Value | Unit | Reference/equation |
|---|---|---|---|---|
| The specific heat capacity | $C_P$ | 820 | J/kg·K | [1] |
| Thermal diffusivity | $\alpha$ | $2.3\cdot 10^{-7}$ | $m^2/s$ | [1] |
| Thermal conductivity | $\lambda$ | 0.792 | W/m·K | $\lambda=\rho\alpha C_p$ |
| Density | $\rho$ | $4.29\cdot 10^3$ | $Kg/m^3$ | [2,3] |
| Electrical conductivity | $\sigma$ | $10^{-2}\div 10^{-4}$ | S/m | [4–8] |

Using the ohmic conductance condition for the spreading resistance in the semispherical geometry, $R_{spr}=\frac{1}{2\pi a\sigma}$, we express the full current as $I=\frac{U}{R_{spr}}=2\pi Ua\sigma$, and substituting for it into Eq. (S 2.3), we obtain: $\omega(a=10nm)=\frac{U^2\sigma_{e^-}(r)}{a^2}\approx 10^{14}W\cdot m^{-3}$. Then the maximal rate of the local temperature growth (without heat loss) in LMO is: $\frac{\partial T}{\partial t}=\frac{\omega}{\rho C_p}=3\cdot 10^7$ K/sec. The cooling rate can be estimated from the heat equation by using the temperature distribution $T(r)$ at a moment $t_0$. The result is: $\frac{\partial T}{\partial t}=\alpha\nabla^2 T=-\frac{\alpha}{\lambda}\frac{\sigma U^2}{r^2}\approx -3\cdot 10^5$ K/sec. *That is, the local temperature rises and decreases almost instantaneously, therefore, we can consider a stationary problem.*

The temperature distribution under the probe in the case of a uniform media can be found as a solution of the heat equation:

$$\frac{\partial T}{\partial t}=\alpha\nabla^2 T. \quad \text{(S 2.4)}$$

In the stationary case $\frac{\partial T}{\partial t}=0$, with a uniform media ($\sigma_{e^-}(r)=\sigma=const$), the temperature gradient provides the balance between the generated and dissipated heat, and the balance equation for the heat flux can de written:

$\lambda \nabla T 2\pi r^2 = -\frac{I^2}{2\cdot 4\pi^2}\int_{r_0}^{r}\frac{4\pi r^2}{r^4\sigma(r)}dr \approx -\frac{I^2}{2\pi\sigma}\int_{r_0}^{r}\frac{dr}{r^2} = -\frac{I^2}{2\pi\sigma}\left(\frac{1}{r_0}-\frac{1}{r}\right)$. Here we can express the temperature gradient as: $\nabla T = -\frac{I^2}{4\pi^2\sigma\lambda r^2}\left(\frac{1}{r_0}-\frac{1}{r}\right)$. Then, after integration we obtain a relation for the temperature distribution in a uniform media for the semispherical case:

$$T(r) = \frac{I^2}{4\pi^2\sigma\lambda}\left(\frac{1}{r_0 r}-\frac{1}{2r^2}\right)+T_\infty \qquad \text{(S 2.5)}$$

Under the tip at $r = r_0$, it takes the form:

$$T(r_0) = \frac{I^2}{8\pi^2\sigma\lambda r_0^2}+T_\infty, \qquad \text{(S 2.6)}$$

where $T_\infty$ is the temperature of the environment (~300 K). For the used geometry (Fig. S2.1.) and size of the point contact (*r0* = 10nm), the critical current for LMO can be estimated from Eq. (S 2.6) as ~2.5-25 nA. At such current, the temperature increase under the probe does not exceed 10 K.

Using the ohmic conductance condition, $R_s = \frac{1}{2\pi r_0\sigma}$, we can express the current as $I = \frac{U}{R_s} = 2\pi U r_0\sigma$, then Eq. (S 2.6) converts to:

$$T(r_0) = \frac{U^2\sigma}{2\lambda}+T_\infty, \qquad \text{(S 2.7)}$$

Hence, the temperature increase under the probe is $\Delta T(r_0) = \frac{U^2\sigma}{2\lambda}$. If we limit the temperature raise, e. g., by setting $\Delta T(r_0) = 10$ K, the maximal voltage can be estimated as $U_{\max} = \sqrt{10\frac{2\lambda}{\sigma}} \approx 40-400$ V.

*The steady-state temperature under the probe, at* $r = r_0$*, is a quadratic function of the bias voltage and does not depend on the tip radius for a given bias voltage because the current is limited by the spreading resistance of the tip-sample contact.* However, in the case of space-charge limited current when $I \sim U^\alpha$, where $\alpha \geq 2$, the temperature increase at $r = r_0$ can depend on the tip radius. *Accounting for the heat dissipation through the metallic probe will result in a decrease of the temperature with an increase of the tip radius.*

*The governing parameter in Eq.* (S 2.7) *is the ratio of the electric conductivity and the thermal conductivity,* $\frac{\sigma}{\lambda}$. In semiconducting materials, the thermal conductivity grows faster with temperature than the electric conductivity, and in a broad class of materials, the ratio $\frac{\lambda}{\sigma}$ can be described by Wiedemann-Franz law: $\frac{\lambda}{\sigma} = LT$, where $L = 2.44 \cdot 10^{-8} W\Omega K^{-2}$ for metals. *Due to the faster growth of thermal conductivity the system is thermostable against the Joule heating.*

Further we consider a simplified model of the probe sample contact with two regions: surface layer (region 1) and bulk (region 2) as shown in Fig. S 2.2.

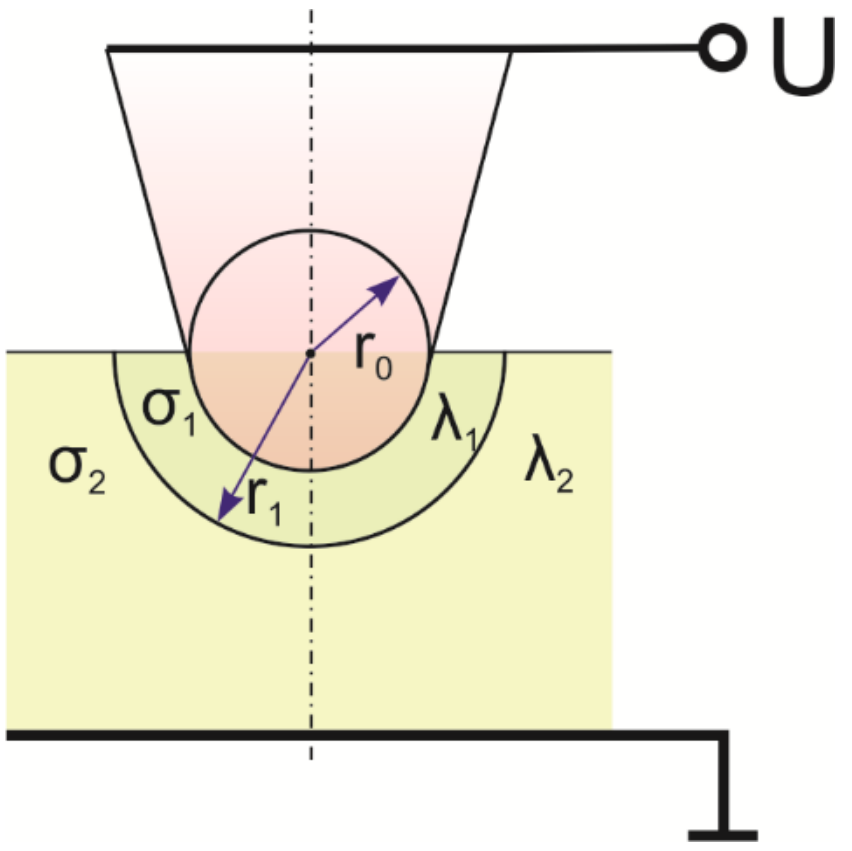

Figure. S 2.2. Schematic of the semispherical geometry for the heat calculations with two regions: the surface layer region with electric and heat conductivities $\sigma_1$, $\lambda_1$, and the bulk region with electric and heat conductivities $\sigma_2$, $\lambda_2$.

With the ohmic conductance, we can write a heat balance equation for region 2 as:

$\lambda_2 2\pi r^2 \nabla T\big|_{r\geq r_1} = -\frac{I^2}{2\cdot 4\pi^2}\left(\int_{r_0}^{r_1}\frac{4\pi r^2}{r^4\sigma_1(r)}dr + \int_{r_1}^{r}\frac{4\pi r^2}{r^4\sigma_2(r)}dr\right)$. After integration, we express the temperature gradient: $\nabla T\big|_{r\geq r_1} = -\frac{I^2}{4\pi^2\lambda_2 r^2}\left(\frac{1}{r_0\sigma_1} - \frac{1}{r_1\sigma_1} + \frac{1}{r_1\sigma_2} - \frac{1}{r\sigma_2}\right)$. The subsequent integration yields the temperature distribution in region 2 as a function of the current *I*:

$$T(r)\big|_{r\geq r_1} = \frac{I^2}{4\pi^2\lambda_2\sigma_1 r r_0}\left(1 - \frac{r_0}{r_1}\left(1 - \frac{\sigma_1}{\sigma_2}\left(1 - \frac{r_1}{2r}\right)\right)\right) + T_\infty. \qquad \text{(S 2.8)}$$

The current $I$ in Eq. (S 2.8) can be expressed as a ratio of the bias voltage over the ohmic resistance $R_s$ of the point contact: $I = \dfrac{U}{R_s}$. The point contact resistance in the semispherical geometry of the model in Fig. S 2.2 is: $R_s = \dfrac{1}{2\pi}\left(\int_{r_0}^{r_1}\dfrac{dr}{r^2\sigma_1} + \int_{r_1}^{\infty}\dfrac{dr}{r^2\sigma_2}\right)$. The resulting expression for $R_s$ is:

$$R_s = \frac{1}{2\pi r\sigma_1 r_0}\left(1 - \frac{r_0}{r_1}\left(1 - \frac{\sigma_1}{\sigma_2}\right)\right). \qquad \text{(S 2.9)}$$

Then, the expression for the current is:

$$I = \frac{U}{R_s} = \frac{2\pi r_0\sigma_1 U}{1 - \dfrac{r_0}{r_1}\left(1 - \dfrac{\sigma_1}{\sigma_2}\right)}. \qquad \text{(S 2.10)}$$

Substituting Eq. (S 2.10) into Eq. (S 2.8), we obtain the temperature distribution in region 2 as a function of the voltage $U$:

$$T(r)\big|_{r\geq r_1} = \frac{\sigma_1 U^2}{\lambda_2}\frac{r_0}{r}\frac{\left(1 - \dfrac{r_0}{r_1}\left(1 - \dfrac{\sigma_1}{\sigma_2}\left(1 - \dfrac{r_1}{2r}\right)\right)\right)}{\left(1 - \dfrac{r_0}{r_1}\left(1 - \dfrac{\sigma_1}{\sigma_2}\right)\right)^2} + T_\infty, \qquad \text{(S 2.11)}$$

According to Eq. (S 2.5) and Eq. (S 2.10), the expression for the temperature distribution in region 1 can be written as:

$$T(r)\big|_{r_0\leq r\leq r_1} = \frac{\sigma_1 U^2}{\lambda_1}\frac{r_0}{r}\frac{\left(1 - \dfrac{r_0}{2r}\right)}{\left(1 - \dfrac{r_0}{r_1}\left(1 - \dfrac{\sigma_1}{\sigma_2}\right)\right)^2} + T_1, \qquad \text{(S 2.12)}$$

where the constant of integration $T_1$ can be found from the continuity condition at $r = r_1$: $T(r_1) = \dfrac{\sigma_1 U^2}{\lambda_1}\dfrac{r_0}{r_1}\dfrac{\left(1 - \dfrac{r_0}{2r_1}\right)}{\left(1 - \dfrac{r_0}{r_1}\left(1 - \dfrac{\sigma_1}{\sigma_2}\right)\right)^2} + T_1$. From Eq. (S 2.11),

$$T(r_1) = \frac{\sigma_1 U^2}{\lambda_2}\frac{r_0}{r_1}\frac{\left(1-\frac{r_0}{r_1}\left(1-\frac{\sigma_1}{2\sigma_2}\right)\right)}{\left(1-\frac{r_0}{r_1}\left(1-\frac{\sigma_1}{\sigma_2}\right)\right)^2} + T_\infty$$, then:

$$T_1 = \frac{\sigma_1 U^2}{\lambda_2}\frac{r_0}{r_1}\frac{\left(1-\frac{r_0}{r_1}\left(1-\frac{\sigma_1}{2\sigma_2}\right)\right)}{\left(1-\frac{r_0}{r_1}\left(1-\frac{\sigma_1}{\sigma_2}\right)\right)^2} - \frac{\sigma_1 U^2}{\lambda_1}\frac{r_0}{r_1}\frac{\left(1-\frac{r_0}{2r_1}\right)}{\left(1-\frac{r_0}{r_1}\left(1-\frac{\sigma_1}{\sigma_2}\right)\right)^2} + T_\infty$$. Substituting it into Eq. (S 2.12), we obtain the temperature distribution at $r_0 \le r \le r_1$:

$$T(r)\Big|_{r_0 \le r \le r_1} = \frac{\sigma_1 U^2}{\lambda_1}\frac{r_0}{r}\frac{\left(1-\frac{r_0}{2r}\right)+\frac{r}{r_1}\left(\frac{\lambda_1}{\lambda_2}\left(1-\frac{r_0}{r_1}\left(1-\frac{\sigma_1}{2\sigma_2}\right)\right)-\left(1-\frac{r_0}{2r_1}\right)\right)}{\left(1-\frac{r_0}{r_1}\left(1-\frac{\sigma_1}{\sigma_2}\right)\right)^2} + T_\infty . \quad \text{(S 2.13)}$$

Under the probe at $r = r_0$, it takes the form:

$$T(r_0) = \frac{\sigma_1 U^2}{\lambda_1}\frac{\frac{1}{2}+\frac{r_0}{r_1}\left(\frac{r_0}{2r_1}-1+\frac{\lambda_1}{\lambda_2}\left(1-\frac{r_0}{r_1}\left(1-\frac{\sigma_1}{2\sigma_2}\right)\right)\right)}{\left(1-\frac{r_0}{r_1}\left(1-\frac{\sigma_1}{\sigma_2}\right)\right)^2} + T_\infty . \quad \text{(S 2.14)}$$

Let us inspect the limiting cases:

1. With high electric and heat conductivities of region 1, $\frac{\lambda_2}{\lambda_1} \to 0$, $\frac{\sigma_2}{\sigma_1} \to 0$, the relation (S 2.14) results in: $T(r_0) = \frac{\sigma_2 U^2}{2\lambda_2} + T_\infty$.

2. In the case of a highly resistive surface region 1, with low electric and heat conductivities, $\frac{\lambda_2}{\lambda_1} \to \infty$, $\frac{\sigma_2}{\sigma_1} \to \infty$, and we obtain: $T(r_0) = \frac{\sigma_1 U^2}{2\lambda_1} + T_\infty$.

The obtained expressions are similar to the expression for a uniform media Eq. (S 2.7).

## Conclusions:

- The temperature under the probe for the case with two regions (Fig. S 2.2.) is defined by the region with the lowest electric and heat conductivity. In our experiments, it is the surface region 1 (Fig. S 2.2.).

- Due to the faster growth of thermal conductivity with temperature, the system is thermostable against the Joule heating. Due to the heat dissipation through the probe, the temperature under the probe will decrease with increasing tip radius.
- The conductivity in $LiMn_2O_4$ is thermally activated and increases with temperature. Therefore, the asymmetry of I-V curves in respect to the voltage bias polarity can be enhanced due to the Joule heating.
- According to the temperature evaluations for the ohmic conduction in a uniform media, the temperature change under the probe in LMO can be neglected if the bias voltages does not exceed 40 V.

---

## Supplementary 3.

## Screening length estimation with account of the steric effect

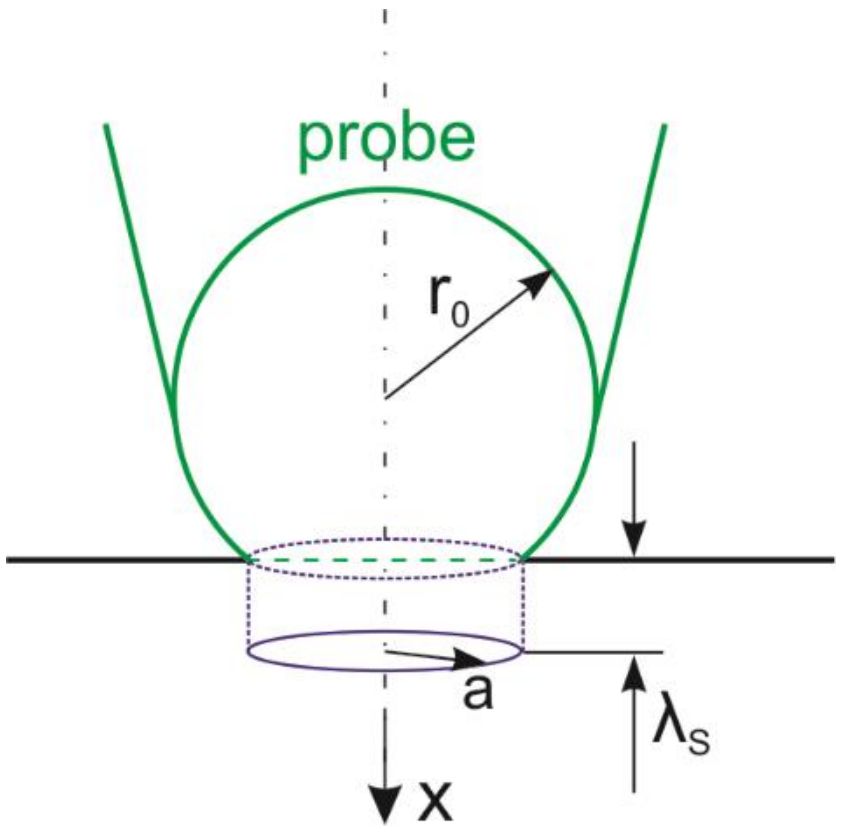

Figure. S3.1. Schematic of a simplified screening charge distribution under a tip apex within a sample surface layer of thickness $\lambda_S \ll a$; $a$ is the tip-sample contact radius.

In the absence of charge injection or/and electric current, the screening length $\lambda_S$ in LMO in the presence of the steric effect can be estimated in the configuration of a planar metallic surface in contact with the LMO. This approach should be used at high electric fields $\frac{eEd}{k_B T} \geq 1$, where $E$ is the electric field in the material and $d$ is the LMO lattice parameter ($d \approx 8$ Å, $E \approx 10^8$ V/m [9]), when the volume ion density in the LMO near the interface reaches the maximal value allowed by the stoichiometry, $n_{\max}$. We assume that the screening charge in LMO is fully confined in a layer of thickness $\lambda_S$ (Fig. S 3.1). The surface charge densities on the metal surface and the screening charge density in the LMO are both equal to $\sigma_S = en_{\max}\lambda_S$ and opposite in sign. To calculate $\lambda_S$, the voltage drop through the double layer can be set equal to the applied voltage bias *U*. The voltage drop across the double layer can be found as a sum of potential differences induced by the charges on the metal surface and in LMO. It is obvious that the potentials induced by the charge in the LMO are equal on both sides of the charged layer, and the LMO contribution to the potential drop is zero. The potential drop due to the charges on the metallic surface is:

$$E_{plane}\lambda_S = \frac{en_{\max}\lambda_S}{2\varepsilon\varepsilon_0}\lambda_S = U .$$

From the second equity we obtain:

$$\lambda_S = \sqrt{\frac{2\varepsilon\varepsilon_0 U}{en_{\max}}} . \qquad \text{(S 3.1)}$$

For LMO's $n_{\max}$ = $1.4\cdot 10^{28}$ $m^{-3}$, $\varepsilon$ = 8.43, and *U* = 10 V (see Table S 5.), $\lambda_S = 0.28$ nm.

We note that $\lambda_S$ in Eq. (S 3.1) coincides with the width of the saturated layer $l^*$, introduced in Ref. [10] based on the solution of the generalized Poisson-Boltzmann equation that takes into account the steric effect. The coincidence take place if the potential drop through the saturated layer is equated to *U*. The full potential drop in the exact solution of Ref. [10] is about 15-20% larger, while the full width of the double layer is by a factor of about two larger.

The screening length under the probe is close to that estimated in the planar geometry because the contact radius $a >> \lambda_S$.

## Supplementary 4.

Table. S4 (Elastic modulus and work function values for different substrate and probe materials used in the experiments)

| Material | Elastic modulus, GPa | Reference | Work Function, eV | Reference |
|---|---|---|---|---|
| $\lambda$-$MnO_2$÷$LiMn_2O_4$ | ~100 | [11,12] | 8-6.2 | [13,14] |
| $W_2C$ | 428 | [14] | 4.58 (1800ºK) | [14] |
| TiN | ~256 | [14] | 4.5 | [15] |
| TiC | ~460 | [14] | 2.6-2.7, 3.8, 4.1 | [16–18] |
| Pt | 290 | [19] | 5.3-6.5 | [20] |
| HOPG | | | 4.7, 4.5 | [21] |

## Supplementary 5.

Table. S5. Physical properties of $LiMn_2O_4$ at room temperature.

| Property | Symbol | Value | Unit | Reference/equation |
|---|---|---|---|---|
| Li-ion concentration for the fully discharged cathode | $n_{\max}$ | $1.4 \cdot 10^{28}$ | $m^{-3}$ | [22,23] |
| Vegard coefficient of chemical expansion | $\beta$ | $\sim 2 \cdot 10^{-30}$ | $m^3$ | [23–27] |
| Poisson coefficient | $\nu$ | ~ 0.3 | | [22,27–29] |
| Electron mobility | $\mu_{e^-}$ | $10^{-11} \div 10^{-13}$ | $m^2V^{-1}s^{-1}$ | $\mu_{e^-} = \frac{\sigma(x)}{q \cdot n_{e^-}(x,t)}$ |
| Maxwell relaxation time | $\tau_M$ | $\sim 7 \times \left(10^{-9} \div 10^{-7}\right)$ | sec | $\tau_M = \frac{\varepsilon\varepsilon_0}{\sigma}$ |
| Electrical conductivity | $\sigma$ | $10^{-2} \div 10^{-4}$ | S/m | [4–8] |
| Permittivity – $\lambda$-$MnO_2 \div LiMn_2O_4$ | $\varepsilon$ | $4.19 \div 8.43$ | | [30,31] |
| Band gap | $E_g$ | 1.43, 1.99, 2.12, 2.42 | eV | [13,31–33] |

## Supplementary 6.

## Poole–Frenkel conduction of a point contact in the semispherical geometry

Poole–Frenkel relation for conductivity is [34,35]:

$$\sigma = \sigma_0 \exp\left(\frac{\beta_{pF}}{2k_B T}\sqrt{E}\right), \tag{S 6.1}$$

where $k_B$ is the Boltzmann constant, $\beta_{pF}$ is the Poole–Frenkel constant $\beta_{pF} = \sqrt{\frac{q^3}{\pi\varepsilon\varepsilon_0}}$ [36] (for LMO $\beta_{pF} \approx 4.3 \cdot 10^{-24}$ J·(m/V)$^{1/2}$), where $q$ is the elementary charge, $\varepsilon_0$ is the permittivity of free space, $\varepsilon$ is the relative dielectric constant and $\sigma_0$ is the low-field conductivity [37–39]. In the electric field of the probe, we use Eq. (S 6.1) for conductivity as a local characteristic in the approximation of a uniform electric field at a small length scale, i.e., at a scale less than the contact radius (~ 6 nm). In the semispherical geometry, we can express the resistance $R$ of the probe-sample point contact as:

$$R = \frac{1}{2\pi}\int_a^\infty \frac{dr}{r^2\sigma(r)} = \int_a^\infty \frac{E(r)dr}{I}, \tag{S 6.2}$$

Using Ohm's law, the electric field in the semispherical geometry can be expressed as:

$$E = \frac{I}{2\pi r^2 \sigma(r)}, \tag{S 6.3}$$

then $r = \sqrt{\frac{I}{2\pi E \sigma(r)}}$, where current *I* is a function of the bias voltage, i. e., $I(U)$. Using the Poole–Frenkel relation for conductivity Eq. (S 6.1), we express the radius as a function of the electric field $r(E)$ at a fixed current $I(U)$ at a fixed bias voltage:

$$r = \sqrt{\frac{I}{2\pi\sigma_0}}\frac{1}{\sqrt{E}}\exp\left(-\frac{\beta_{pF}\sqrt{E}}{4k_B T}\right). \tag{S 6.4}$$

In the last expression, the radius is expressed as a function of the local value of the electric field *E*. From Eq. (S 6.4), the differential $dr = -\frac{dE}{E}\sqrt{\frac{I}{2\pi\sigma_0}}\left(\frac{1}{2\sqrt{E}} + \frac{\beta}{4}\right)\exp\left(-\frac{\beta\sqrt{E}}{2}\right)$, where we use a short notation $\beta = \frac{\beta_{pF}}{2k_B T}$. Substitution for *r* (Eq. (S 6.4)) and d*r* in Eq. (S 6.2) yields:

$R = -\frac{1}{\sqrt{2\pi I\sigma_0}}\int_{E(a)}^{E(\infty)}\left(\frac{1}{2\sqrt{E}} + \frac{\beta}{4}\right)\exp\left(-\frac{\beta\sqrt{E}}{2}\right)dE$. After integration, we obtain:

$$R = \frac{4}{\beta\sqrt{2\pi I\sigma_0}}\left(1 - \left(1 + \frac{\beta\sqrt{E(a)}}{4}\right)\exp\left(-\frac{\beta\sqrt{E(a)}}{2}\right)\right). \tag{S 6.5}$$

Using relation $I = \frac{U}{R}$, we express the current as a function of electric field at the contact ($E(a)$):

$$I = \frac{\pi\sigma_0\left(\frac{\beta_{pF}}{2k_B T}\right)^2 U^2 \exp\left(\frac{\beta_{pF}}{2k_B T}\sqrt{E(a)}\right)}{8\left(\exp\left(\frac{\beta_{pF}}{4k_B T}\sqrt{E(a)}\right) - 1 - \frac{\beta_{pF}}{8k_B T}\sqrt{E(a)}\right)^2}. \tag{S 6.6}$$

Expressing current at $r = a$ from Eq. (S 6.4) as $I = E(a) 2\pi a^2 \sigma_0 \exp\left(\frac{\beta_{pF}\sqrt{E(a)}}{2k_B T}\right)$, and setting it equal to the current in Eq. (S 6.6), we can find the electric field $E(a)$ near the tip as an implicit function of the bias voltage:

$$\frac{8k_B T\sqrt{E(a)}}{\beta_{pF}}\left(\exp\left(\frac{\beta_{pF}}{4k_B T}\sqrt{E(a)}\right) - 1 - \frac{\beta_{pF}}{8k_B T}\sqrt{E(a)}\right) = \frac{U}{a}. \tag{S 6.7}$$

For a low power index in the exponent $\frac{\beta_{pF}}{4k_B T}\sqrt{E(a)} \le 1$, at $E(a) \le 1.5\cdot 10^7$ V/m, Eqs. (S 6.6) and (S 6.7) can be reduced to:

$$I \approx 2\pi\sigma_0 U^2 \frac{\exp\left(\frac{\beta_{pF}}{2k_B T}\sqrt{E(a)}\right)}{E(a)}, \tag{S 6.8}$$

and (see Appendix (A 6.5)) to:

$$E(a) \approx \frac{U}{a\left(1 + \frac{\beta_{pF}}{4k_B T}\sqrt{\frac{U}{a}}\right)}, \tag{S 6.9}$$

correspondingly. Then, the resulting approximation for $I(U)$ is:

$$I \approx 2\pi a\sigma_0 U\left(1 + \frac{\beta_{pF}}{4k_B T}\sqrt{\frac{U}{a}}\right)\exp\left(\frac{\beta_{pF}}{2k_B T}\sqrt{\frac{U}{a\left(1 + \frac{\beta_{pF}}{4k_B T}\sqrt{\frac{U}{a}}\right)}}\right) \tag{S 6.10}$$

## Appendix:

After power series expansion of the exponent to the second order, the equation $\frac{8k_B T\sqrt{E(a)}}{\beta_{pF}}\left(\exp\left(\frac{\beta_{pF}}{4k_B T}\sqrt{E(a)}\right) - 1 - \frac{\beta_{pF}}{8k_B T}\sqrt{E(a)}\right) = \frac{U}{a}$ is reduced to an approximate form:

$$E(a)\left(1 + \frac{\beta_{pF}}{4k_B T}\sqrt{E(a)}\right) = \frac{U}{a}. \tag{A 6.1}$$

To find solution for $E(a)$ of Eq. (A 6.1), we consider an equation of the form:

$$x^2(1 + \alpha x) = c. \tag{A 6.2}$$

Where $x = \sqrt{E(a)}$, $\alpha = \frac{\beta_{pF}}{4k_B T}$ and $c = \frac{U}{a}$. The positive solution of Eq. (A 6.2) at $x \geq 0$, $\alpha \geq 0$, and $c \geq 0$ is an iterative relation:

$$x = \sqrt{\frac{c}{1+\alpha\sqrt{\frac{c}{1+\alpha\sqrt{\frac{c}{\ddots}}}}}}. \qquad \text{(A 6.3)}$$

The solution A 6.3 can be used to express an approximate solution of Eq. (A 6.2) with a required degree of accuracy taking *c* satisfying the inequity $\alpha\sqrt{\frac{c}{2}} \leq 1$ as an initial value:

$$x = \sqrt{\frac{c}{1+\alpha\sqrt{\frac{c}{1+\alpha\sqrt{\ddots \frac{c}{1+\alpha\sqrt{c}}}}}}}. \qquad \text{(A 6.4)}$$

With $\frac{\beta_{pF}}{4k_B T}\sqrt{E(a)} < 1$, we used the second iteration in (A 6.4) and obtained an approximate solution for Eq. (A 6.1):

$$E(a) \approx \frac{U}{a\left(1+\frac{\beta_{pF}}{4k_B T}\sqrt{\frac{U}{a}}\right)}. \qquad \text{(A 6.5)}$$

---

## Supplementary 7

## Contact radii at different loading forces.

Table S7.

| Set Point, nA | Loading, *L*, nN | Contact radius, *a*, nm |
|---|---|---|
| 0,7 | 154 | 3.26 |
| 2 | 440 | 4.63 |
| 3 | 660 | 5.3 |
| 4 | 880 | 5.8 |

Were radius of the Hertzian contact is given by a relation:

$$a = \sqrt[3]{\frac{3R_{tip}L}{4E_{eff}}}, \qquad \text{(S 7.1)}$$

where $L$ is the static load applied to the tip and $R_{tip}$ the curvature radius of the tip. $E_{eff}$ is defined as $\frac{1}{E_{eff}} = \frac{1-\nu_S^2}{E_S} + \frac{1-\nu_{tip}^2}{E_{tip}}$, where $E_S$ and $E_{tip}$ are respectively the elastic moduli of the sample and the tip and $\nu_S$ and $\nu_{tip}$ are respectively the Poisson ratios of the sample and the tip. In the case when $E_S << E_{tip}$ , $E_{eff} \approx E_S$. Data on the elastic moduli and work functions for materials of the probe and $LiMn_2O_4$ are listed in the Table S4.

## Supplementary 8.

## Calculations of the second and third derivatives, $\frac{\partial^2 I}{\partial U^2}$, $\frac{\partial^3 I}{\partial U^3}$, from the inverse function $U(I) = I^{\frac{1}{\alpha}}\tilde{R}_s + I\tilde{R}_b$.

The second derivative of an I-V curve can be used for extraction of the relation between $\tilde{R}_b$ and $\tilde{R}_s$. The observed second derivative extremum in Fig. 6(a) can be identified analytically from the relation $\frac{\partial^3 I}{\partial U^3} = 0$.

Calculation of $\frac{\partial^3 I}{\partial U^3}$ from the equation $U(I) = I^{\frac{1}{\alpha}}\tilde{R}_s + I\tilde{R}_b$ is the following:

1th derivative: $\frac{\partial I}{\partial U} = \frac{1}{\frac{\partial U}{\partial I}} = \frac{1}{\frac{1}{\alpha}\tilde{R}_s I^{\left(\frac{1-\alpha}{\alpha}\right)} + \tilde{R}_b}$,

2th derivative: $\frac{\partial}{\partial U}\left(\frac{\partial I}{\partial U}\right) = \frac{-\frac{1-\alpha}{\alpha^2}\tilde{R}_s I^{\left(\frac{1-2\alpha}{\alpha}\right)}}{\left(\frac{1}{\alpha}\tilde{R}_s I^{\left(\frac{1-\alpha}{\alpha}\right)} + \tilde{R}_b\right)^3}$,

3th derivative: $\frac{\partial}{\partial U}\left(\frac{\partial^2 I}{\partial U^2}\right) = \frac{3\frac{(1-\alpha)^2}{\alpha^4}\tilde{R}_s^2 I^{2\left(\frac{1-2\alpha}{\alpha}\right)}}{\left(\frac{1}{\alpha}\tilde{R}_s I^{\left(\frac{1-\alpha}{\alpha}\right)} + \tilde{R}_b\right)^5} - \frac{\frac{(1-\alpha)(1-2\alpha)}{\alpha^3}\tilde{R}_s I^{\left(\frac{1-3\alpha}{\alpha}\right)}}{\left(\frac{1}{\alpha}\tilde{R}_s I^{\left(\frac{1-\alpha}{\alpha}\right)} + \tilde{R}_b\right)^4}$,

From the relation $\frac{\partial^3 I}{\partial U^3} = 0$, we find:

$$\tilde{R}_b = \frac{\alpha - 2}{\alpha(2\alpha - 1)} \tilde{R}_s I_p^{\left(\frac{1-\alpha}{\alpha}\right)} . \qquad \text{(S 8.1)}$$

Where $I_p = I(U_{peak})$ is the current value at the peak of the second derivative $\frac{\partial^2 I}{\partial U^2}$ (see Fig. 6(a) in the main text of the paper). The impedance at the maximum peak of the 2nd derivative is a function of the power $\alpha$. From the relation (S 8.1), $I_p \tilde{R}_b = \frac{\alpha - 2}{\alpha(2\alpha - 1)} I_p^{\frac{1}{\alpha}} \tilde{R}_s$ and:

$$\frac{I_p^{\frac{1}{\alpha}} \tilde{R}_s}{I_p \tilde{R}_b} = \frac{\alpha(2\alpha - 1)}{\alpha - 2} . \qquad \text{(S 8.2)}$$

The position of the peak over the bias voltage can be calculated as the following:

Expressing $I_p R_b$ from Eq. (S 8.2) and using $U(I) = I^{\frac{1}{\alpha}} \tilde{R}_s + I\tilde{R}_b$, we can write:

$$U(I_p) = I_p^{\frac{1}{\alpha}} \tilde{R}_s \left( 1 + \frac{\alpha - 2}{\alpha(2\alpha - 1)} \right) . \qquad \text{(S 8.3)}$$

Expressing $I_p^{\frac{1}{\alpha}}$ from Eq. (S 8.2) and substituting into Eq. (S 8.3), we obtain the voltage peak position $U_p$ as function of three parameters, $\tilde{R}_b$, $\tilde{R}_s$, and $\alpha$:

$$U_p = \left( \frac{(\alpha - 2)}{\alpha(2\alpha - 1)} \frac{\tilde{R}_s^{\alpha}}{\tilde{R}_b} \right)^{\frac{1}{\alpha - 1}} \frac{2\alpha^2 - 2}{\alpha(2\alpha - 1)} . \qquad \text{(S 8.4)}$$

## Supplementary 9.
## Space charge limited current (SCLC) in the one-dimensional and semispherical cases

### 1. SCLC in the one-dimensional case

For so-called lifetime semiconductors [39], i. e., semiconductors with a carrier lifetime $\tau_0$ greater than the Maxwell relaxation time, $\tau_0 > \tau_M$, the stationary, time-independent, SCLC theory relies on simultaneous solution of the Poisson and charge continuity equations [39–41]. The Poisson equation for the SCLC in the one-dimensional case is:

$$\frac{\partial E}{\partial x} = \frac{q(\rho_f + \rho_t)}{\varepsilon\varepsilon_0}, \qquad \text{(S 9.1)}$$

where $\rho_f$ and $\rho_t$ are, respectively, the densities of injected free and trapped charge carriers. The charge continuity equation $\nabla J = 0$ for the current density $J$ in the one-dimensional case yields $J = const$, and Ohm's law states:

$$J = q\rho_f \mu E, \qquad \text{(S 9.2)}$$

where carrier drift mobilities $\mu - \mu_n$ for electrons and $\mu_p$ for holes–are functions of coordinates:

$$\mu(x, E) = \mu_0 S_\mu(x), \qquad \text{(S 9.3)}$$

where $\mu_0 = \frac{1}{d}\int_0^d \mu(x)dx$ is the average mobility over the sample thickness $d$. Depending on the bias voltage polarity (Fig. 3. (c), (e)), these are electrons or holes. The densities of electrons, *n*, and holes, *p*, are given by [39]:

$$n_t(x) = \int_{E_l}^{E_u} h_n(E_w, x) f_n(E_w) dE_w, \qquad \text{(S 9.4)}$$

$$p_t(x) = \int_{E_l}^{E_u} h_p(E_w, x) f_p(E_w) dE_w, \qquad \text{(S 9.5)}$$

$$n_f(x) = N_c \exp\left(\frac{E_{Fn} - E_c}{k_B T}\right), \qquad \text{(S 9.6)}$$

$$p_f(x) = N_v \exp\left(-\frac{E_{Fp} - E_v}{k_B T}\right). \qquad \text{(S 9.7)}$$

Where $k_B$ is the Boltzmann constant, $E_w$ is the energy, $E_c$ and $E_v$ are (carrier-concentration-dependent) energy levels of the bottom of the conduction band and top of the valence band, respectively, $E_{Fn}$ and $E_{Fp}$ are quasi-Fermi levels for electrons and holes (in equilibrium $E_{Fn} = E_{Fp} = E_F$); $N_c$ and $N_v$ are, respectively, the effective densities of states (number of states per unit volume) in the conduction and valence bands, which are given by:

$$N_c = 2\left(\frac{m_n^* k_B T}{2\pi\hbar^2}\right)^{\frac{3}{2}}, \qquad \text{(S 9.8)}$$

$$N_v = 2\left(\frac{m_p^* k_B T}{2\pi\hbar^2}\right)^{\frac{3}{2}}. \qquad \text{(S 9.9)}$$

where $\hbar$ is the reduced Planck constant, $m_n^*$ and $m_p^*$ are the effective masses of electron and hole, respectively. The distribution function for trap density as a function of the energy $E_w$ and distance $x$ from the injecting contact for electrons and holes can be written as:

$$h_n(E_w, x) = N_{nt}(E_c - E_w)S_n(x), \qquad \text{(S 9.10)}$$

$$h_p(E_w, x) = N_{pt}(E_w - E_v)S_p(x), \qquad \text{(S 9.11)}$$

where $N_t(E_w)$ and $S(x)$ represent, respectively, the energy and spatial distribution functions for electron and/or hole trap densities. The probability that a trap will capture an electron follows the Fermi-Dirac statistics:

$$f_n(E_w) = \frac{1}{1 + g_n^{-1}\exp\left(\frac{E_w - E_{Fn}}{k_B T}\right)}, \qquad \text{(S 9.12)}$$

where $T$ is the temperature. In turn, the probability that a trap will capture a hole is:

$$f_p(E_w) = \frac{1}{1 + g_p \exp\left(\frac{E_{Fp} - E_w}{k_B T}\right)}, \qquad \text{(S 9.13)}$$

where $g_n$ and $g_p$ are the degeneracy factors for the electron and hole traps, respectively.

Following Eqs. (S 9.10) and (S 9.11), when traps are distributed exponentially over energy in the band gap (Fig. S 9.1), the distribution functions for the trap densities take the form:

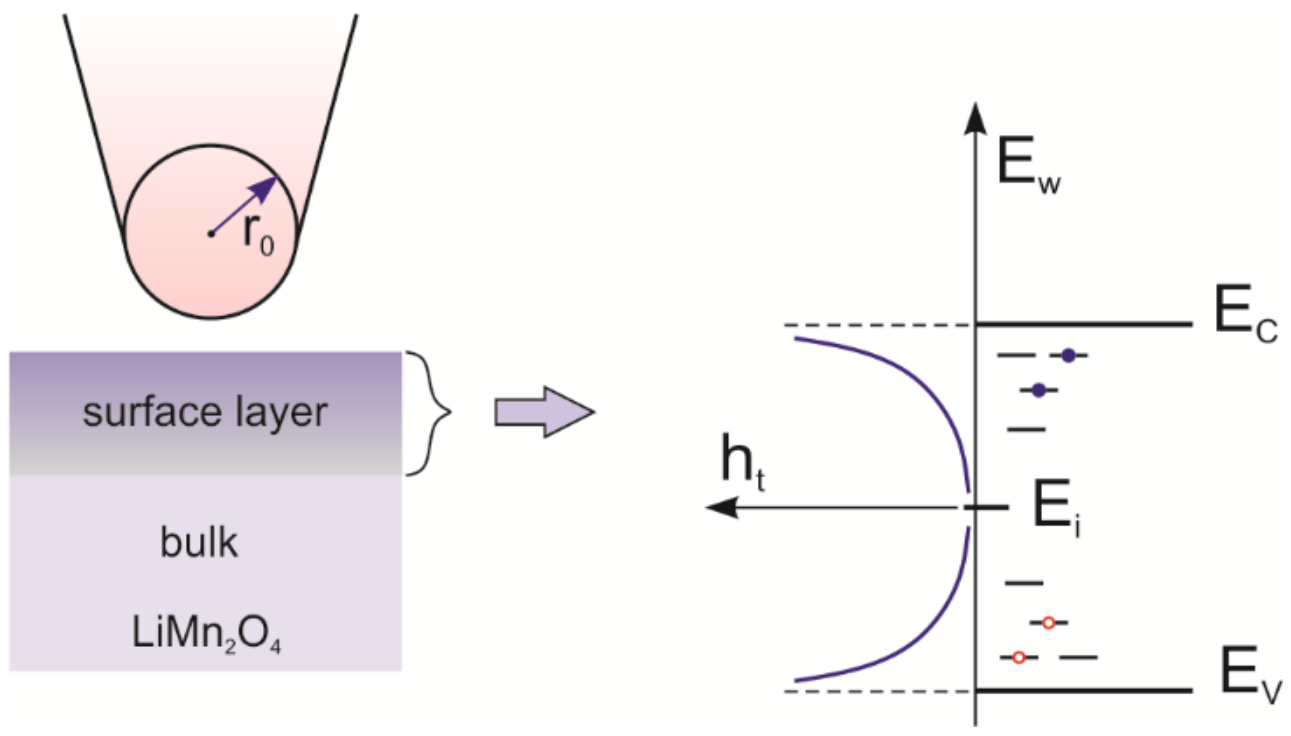

Figure. S 9.1. Left: Schematic of the sample near the surface region with a surface layer filled with charge traps. Right: Schematic of the energy distributions, $h_{tr}(E_w)$, for donor-like ($E_V \le E_w \le E_i$) and acceptor-like ($E_i \le E_w \le E_C$) traps. The traps are exponentially distributed over energy.

for electrons:

$$h_n(E_w,x)=\frac{H_n}{k_BT_{c,n}}\exp\left(\frac{E_w-E_c}{k_BT_{c,n}}\right)S_n(x), \tag{S 9.14}$$

And for holes:

$$h_p(E_w,x)=\frac{H_p}{k_BT_{c,p}}\exp\left(-\frac{E_w-E_v}{k_BT_{c,p}}\right)S_p(x). \tag{S 9.15}$$

Where $H$ is the density of traps, and $T_c$ is a characteristic constant of the distribution. If $T_c > T$, and/or if we take $T=0$, we get: $f_p(E)=1$ for $E_{Fp} < E_w < \infty$ and $f_p(E_w)=0$ for $E_w < E_{Fp}$; $f_n(E_w)=1$ for $E_w < E_{Fn}$ and $f_n(E_w)=0$ for $E_{Fn} < E_w < \infty$. With this, we obtain:

$$\begin{aligned} n_t(x)&=\int_0^{E_{Fn}}\frac{H_n}{k_BT_{c,n}}\exp\left(\frac{E_w-E_c}{k_BT_{c,n}}\right)S_n(x)dE_w = \\ &=H_n\exp\left(\frac{E_{Fn}-E_c}{k_BT_c}\right)S_n(x)= \\ &=H_n\left(\frac{n_f}{N_c}\right)^{\frac{T}{T_{c,n}}}S_n(x) \end{aligned}, \tag{S 9.16}$$

$$\begin{aligned} p_t(x)&=\int_{E_{Fp}}^{\infty}\frac{H_p}{k_BT_{c,p}}\exp\left(-\frac{E_w-E_v}{k_BT_{c,p}}\right)S_p(x)dE_w = \\ &=H_p\exp\left(-\frac{E_{Fp}-E_v}{k_BT_{c,p}}\right)S_p(x)= \\ &=H_p\left(\frac{p_f}{N_v}\right)^{\frac{T}{T_{c,p}}}S_p(x) \end{aligned}. \tag{S 9.17}$$

Equations (S 9.16) and (S 9.17) can be rewritten in short notation:

$$\rho_f = A\left(\frac{\rho_t}{S(x)}\right)^l, \tag{S 9.18}$$

where $l=\frac{T_c}{T}$, $A_n=\frac{N_c}{H_n^{\,l_n}}$ for electrons and $A_p=\frac{N_v}{H_p^{\,l_p}}$ for holes. With $\rho_f << \rho_t$, $\rho_f$ can be neglected in the Poisson equation Eq. (S 9.1), and expressing $\rho_f$ in Eq. (S 9.2) from $J$, Eq. (S 9.1) can be written down as:

$$\frac{\partial E}{\partial x}=\frac{q}{\varepsilon\varepsilon_0}\left(\frac{\rho_f}{A}\right)^{\frac{1}{l}}S(x)=\frac{q}{\varepsilon\varepsilon_0}\left(\frac{J}{Aq\mu E}\right)^{\frac{1}{l}}S(x), \tag{S 9.19}$$

Using Eq. (S 9.3) for mobility, $\mu = \mu_0 S_\mu(x)$, where $S_\mu(x)$ is spatial distribution functions for electron and/or hole mobility, after integration, we obtain $\frac{l}{l+1} E^{\frac{l+1}{l}} = \frac{q^{\frac{l-1}{l}}}{\varepsilon\varepsilon_0}\left(\frac{J}{A\mu_0 E}\right)^{\frac{1}{l}} \int_0^x \frac{S(x')}{S_\mu(x')^{\frac{1}{l}}} dx'$, and then:

$$E = q^{\frac{l-1}{l+1}} \left(\frac{l+1}{l}\frac{1}{\varepsilon\varepsilon_0}\right)^{\frac{l}{l+1}} \left(\frac{J}{A\mu_0}\right)^{\frac{1}{l+1}} \left(\int_0^x \frac{S(x')}{S_\mu(x')^{\frac{1}{l}}} dx'\right)^{\frac{l}{l+1}}. \qquad \text{(S 9.20)}$$

Integration of Eq. (S 9.20) noting that $U = \int_0^d E(x)dx$, where $d$ is the sample thickness, gives:

$$U = q^{\frac{l-1}{l+1}} \left(\frac{l+1}{l}\frac{1}{\varepsilon\varepsilon_0}\right)^{\frac{l}{l+1}} \left(\frac{J}{A\mu_0}\right)^{\frac{1}{l+1}} \frac{l+1}{2l+1} d_{eff}^{\frac{2l+1}{l+1}}, \qquad \text{(S 9.21)}$$

$$J = q^{1-l} \left(\frac{2l+1}{l+1} U\right)^{l+1} \left(\frac{l}{l+1}\varepsilon\varepsilon_0\right)^{l} A\mu_0 \cdot d_{eff}^{-(2l+1)}. \qquad \text{(S 9.22)}$$

where:

$$d_{eff} = \left(\frac{2l+1}{l+1} \int_0^d \left(\int_0^x \frac{S(x')}{S_\mu(x')^{\frac{1}{l}}} dx'\right)^{\frac{l}{l+1}} dx\right)^{\frac{l+1}{2l+1}}. \qquad \text{(S 9.23)}$$

Trap states with energies within a few $k_B T$ near the mobility edge (shallow traps) are characterized with a finite trapping time. After being trapped for a characteristic time $\tau_{tr}$, a trapped polaron can be thermally activated and released into the conduction band. According to the multiple trap-and-release (MTR) model [42], the effective mobility

$$\mu_{eff} = \mu_0 \frac{\tau}{\tau + \tau_{tr}}, \qquad \text{(S 9.24)}$$

where $\tau_{tr}$ is the average trapping time on shallow traps, and $\tau$ is the average time that is spend by a polaron diffusively traveling between consecutive trapping events. In the case $\tau_{tr} >> \tau$, the charge transport is dominated by trapping, and $\mu_{eff} = \mu_0 \frac{\tau}{\tau_{tr}} \propto \exp\left(-\frac{W_{tr}}{k_B T}\right)$. (As a result, the charge mobility in the surface layer with a higher trap density is reduced in comparison with its intrinsic bulk value $\mu_0$).

Equation S 9.24 assumes a direct interdependence between $S(x)$ and $\left(S_\mu(x)\right)^{-1}$ in Eq. (S 9.23). Making a change of variables $\tilde{S}(x)=\frac{S(x)}{S_\mu(x)^{\frac{1}{l}}}$, we rewrite Eq. (S 9.23) as:

$$d_{eff}=\left(\frac{2l+1}{l+1}\int_0^d\left(\int_0^x\tilde{S}(x')dx'\right)^{\frac{l}{l+1}}dx\right)^{\frac{l+1}{2l+1}}. \quad \text{(S 9.25)}$$

The space charge distribution can be obtained for a uniform media with $\tilde{S}(x)=1$ from Eqs. (S 9.20), (S 9.18), and (S 9.2):

$E=\frac{J}{qA\rho_t^{\,l}\mu_0}=q^{\frac{l-1}{l+1}}\left(\frac{l+1}{l}\frac{1}{\varepsilon\varepsilon_0}\right)^{\frac{l}{l+1}}\left(\frac{J}{A\mu_0}\right)^{\frac{1}{l+1}}x^{\frac{l}{l+1}}$, at $\rho_f << \rho_t$ we can put equality between trapped charge and injected charge $\rho_t=\rho_i$, then:

$$\rho_i=q^{-\frac{2}{l+1}}\left(\frac{J\varepsilon\varepsilon_0}{A\mu_0}\frac{l}{l+1}\right)^{\frac{1}{l+1}}x^{-\frac{1}{l+1}}. \quad \text{(S 9.26)}$$

For analysis of SCLC evolution into a spatially separated ohmic conduction (Fig. S 9.2), we use the regional approximation method [41]. The method is based on additive contributions of injected $\rho_i$ and equilibrium $\rho_0$ carriers into the full current density $J=q(\rho_i+\rho_0)\mu E$. The obtained expression for the space charge distribution Eq. (S 9.26) is valid in a uniform media, and both Eq. (S 9.26) and I-V dependence (Eqs. (S 9.21) and (S 9.22)) are valid in the case when the density of the injected charge carriers is higher than the equilibrium charge carrier density, i. e., $\rho_i>\rho_0$, in the space interval $0\le x\le d$. At a lower density of the injected charge, $\rho_i\le\rho_0$, we have the ohmic conduction regime with a conductivity determined by the equilibrium carrier density $\sigma_0=q\mu_0\rho_0$. The size $x_c$ of the space charge region can be estimated if we substitute the equilibrium charge carriers density $\rho_0$ for $\rho_i$ in the left part of Eq. (S 9.26) :

$$x_c=\frac{J\varepsilon\varepsilon_0}{Aq^2\mu_0\rho_0^{\,l+1}}\frac{l}{l+1}=\frac{J\varepsilon\varepsilon_0}{Aq\sigma\rho_0^{\,l}}\frac{l}{l+1}. \quad \text{(S 9.27)}$$

In the model in Fig. S 9.1, the one-dimensional (planar) approach can be applied to the surface layer if its thickness *t* is smaller than the tip radius, $t<r_0$, otherwise we should consider a three-dimensional case. For the SCLC regime, $\rho_i>\rho_0$, the maximal thickness *t* should be less than $x_c$ and can be estimated from Eq. (S 9.27). At equilibrium, the charge carrier concentration $\rho_0\sim\frac{n_{\max}}{2}$

, where $n_{max}$= 1.4·10$^{28}$ m$^{-3}$ is the maximal lithium concentration in the fully discharged $LiMn_2O_4$, estimation for $l$ = 1, $J$ = 10$^{7}$Am$^{-2}$, $\sigma = (2-6)\cdot 10^{-6}$ S/m, yields:

$$t = \frac{J\varepsilon\varepsilon_0}{Aq\sigma\left(\frac{n_{max}}{2}\right)^l}\frac{l}{l+1} = \frac{10^7\cdot 8\cdot 8.85\cdot 10^{-12}}{A\cdot 1.6\cdot 10^{-19}(2-6)10^{-6}0.7\cdot 10^{28}}\frac{1}{2} = \frac{(3.7-11.1)}{A}\cdot 10^{-8}\text{, m}$$

here $A = \frac{N}{H^l}$, where $N$ is effective density of quantum states in conduction/valence band. For semiconductors (Si, Ge, GaAs), $N$ is ~10$^{18}$-10$^{19}$ cm$^{-3}$. $H$ is the trap density. The typical trap density in the bulk for semiconducting materials varies in a range 10$^{10}$-10$^{18}$ cm$^{-3}$.[43–48] In the surface layer (near the probe-sample interface), the trap density can achieve values of 10$^{18}$-10$^{22}$ cm$^{-3}$. Using these data, we can estimate the thicknesses $t$ ~ 10$^{-3}$-10$^{-9}$m. For the parameters given above, an estimate at $l$ = 2 yields values of $t$ ~ 10$^{-9}$-10$^{-17}$ m. These estimates are provided here to support the relevance of the model in Fig. S 9.1, where we assume that the 1D SCLC regime can be realized in the surface layer with a thickness of a few nanometers.

In the case of a hole injection process, the injected holes move in the direction opposite to the electron current. At the boundary between ohmic and SCLC regions, electron-hole recombination takes place. Therefore, for both the cases, the hole injection and electron injection, in the regional approximation method, we can use majority carrier concentration in the calculations. In LMO, the majority carriers are electrons.

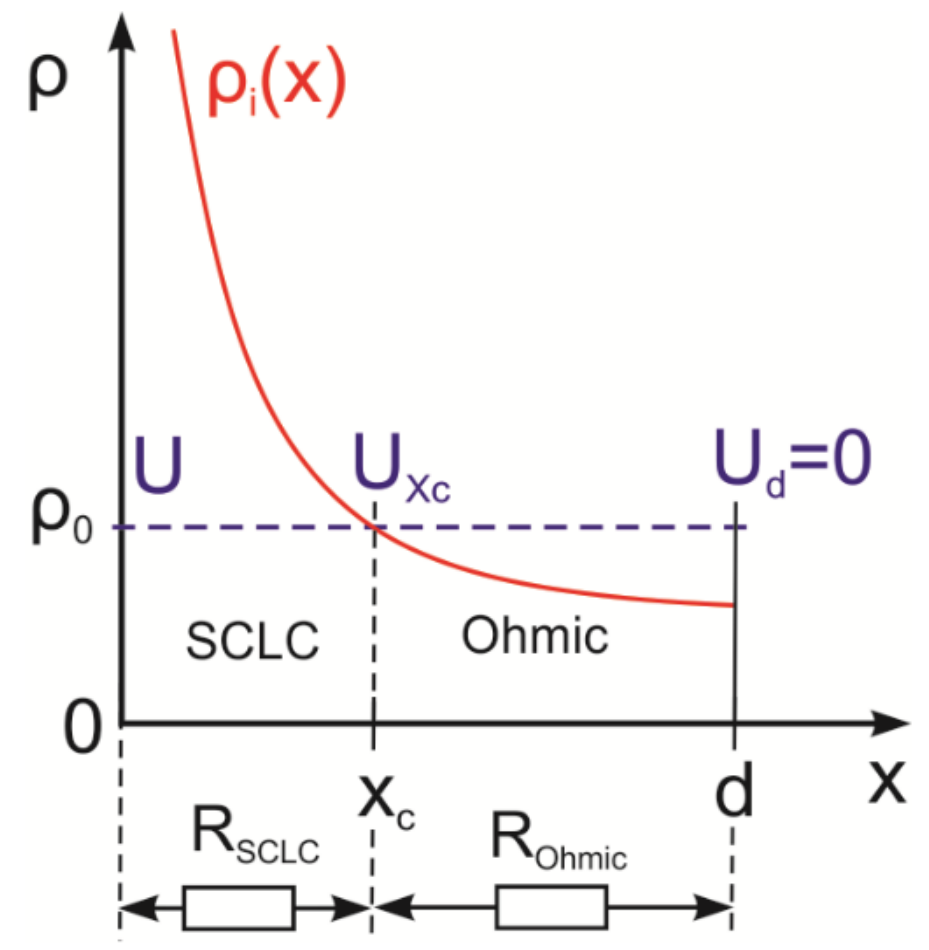

Figure. S 9.2. Illustration of the regional approximation method. Schematic of the equilibrium, $\rho_0$, and the injected, $\rho_i$, charge density distributions along the coordinate. The charge is injected from the left.

The current in the SCLC region $0 \le x \le x_c$, where $\rho_i(x) \ge \rho_0$ (Fig. S 9.2) can be expressed using Eq. (S 9.22) by substitution of $x_c$, Eq. (S 9.27), for $d_{eff}$:

$$J = \sqrt{\frac{2l+1}{l}\frac{q\rho_0}{\varepsilon\varepsilon_0}} Aq\mu_0\rho_0{}^l \sqrt{U - U_{Xc}} \,. \tag{S 9.28}$$

Here $U - U_{Xc}$ is voltage drop across the space charge region (Fig. S 9.2).

The current in the ohmic region, $x_c \le x \le d$, is:

$$J = q\mu_0\rho_0 \frac{U_{Xc}}{d - x_c} \,. \tag{S 9.29}$$

Expressing $U$ from Eqs. S 9.28 and S 9.29 as $U = (R_{SCLC} + R_{Ohmic})J = (U - U_{Xc}) + U_{Xc}$, we obtain:

$$U = \frac{\varepsilon\varepsilon_0\left(1 - A\rho_0{}^{l-1}\frac{2l+1}{l+1}\right)}{q\rho_0\left(Aq\mu_0\rho_0{}^l\right)^2}\frac{l}{2l+1}J^2 + \frac{d}{q\mu_0\rho_0}J \,. \tag{S 9.30}$$

The expression Eq. (S 9.30) is relevant at $U \le U_c$, where the critical voltage $U_c$ can be obtained from Eqs. (S 9.21) and (S 9.27) at $x_c = d$:

$$U_c = \frac{(l+1)^2}{(2l+1)l}\frac{q\rho_0}{\varepsilon\varepsilon_0}d^2 \,. \tag{S 9.31}$$

The relation Eq. (S 9.30) assumes the ohmic I-V dependence $J \propto U$ at small voltages and $J \propto \sqrt{U}$ before transition to SCLC regime. Transition to the SCLC regime is starting at $U > U_c$. The size of the space charge region for the one-dimensional case is a function of conductivity, $x_c \sim \frac{J}{\sigma}$ (Eq. (S 9.27)). Consequently, $x_c$ is two orders of magnitude larger for the highly resistive surface layer than for the bulk, and the space charge can be confined within the surface layer approximately two decades of current values, from ~ 0.01 nA to ~ 1 nA (Fig 7 (b), region 1). When concentration of the injected charge achieves a value equal to or higher than the equilibrium charge carrier concentration in the bulk, the space charge extends into the bulk where a spherically symmetric three-dimensional model can be used for description of the space charge transfer (Fig. S 9.3).

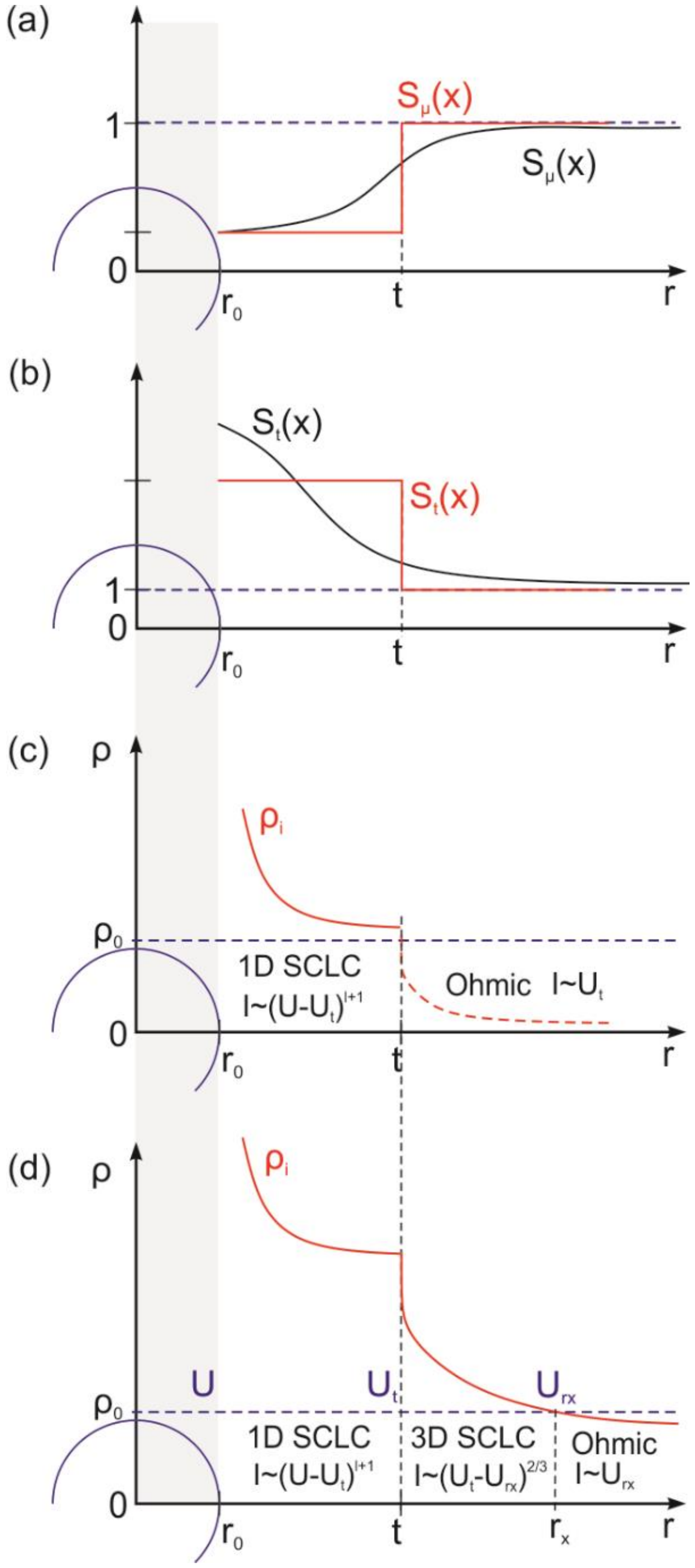

Figure. S 9.3. Schematic of the spatial distribution functions for traps, carrier drift mobility, and injected charge. (a) Black line: spatial distribution function for the carrier drift mobility; red line: its approximation by a step function. (b) Black line: trap spatial distribution function; red line: its approximation by a step function. (c) Injected charge distribution for the one-dimensional SCLC regime in the surface layer. (d) Injected charge distribution for the one-dimensional SCLC regime in the surface layer and three-dimensional SCLC in the bulk. $r_0$ is the tip radius; $t$ is the thickness of the surface layer; $r_x$ is regional boundary coordinate where the injected charge density is equal to the equilibrium charge density, $\rho_i = \rho_0$.

## 2. SCLC in the semispherical approximation

The calculations are performed for a uniform media according to Refs. [39–41]. Poisson's equation for space charge limited current (SCLC) in the spherical geometry is:

$$\frac{1}{r^2}\frac{\partial}{\partial r}\left(r^2E\right)=\frac{q\left(\rho_f+\rho_t\right)}{\varepsilon\varepsilon_0}. \qquad \text{(S 9.32)}$$

For the half-space, the full current can be expressed as:

$$I=J2\pi r^2=\rho_f q\mu_0 E2\pi r^2. \qquad \text{(S 9.33)}$$

All constants are defined according to Eqs. S 8.4-18. For an exponential trap distribution, we have $\rho_f=A\left(\frac{\rho_t}{S(x)}\right)^l$, where $l=\frac{T_c}{T}$, $A_n=\frac{N_c}{H_n^{\,l_n}}$ for electrons and $A_p=\frac{N_v}{H_p^{\,l_p}}$ for holes. At $\rho_f \ll \rho_t$ and $S(x)=1$, combining Eqs. (S 9.32) and (S 9.33), we obtain:

$$\frac{1}{r^2}\frac{\partial}{\partial r}\left(r^2E\right)=\frac{q}{\varepsilon\varepsilon_0}\left(\frac{\rho_f}{A}\right)^{\frac{1}{l}}=\frac{q}{\varepsilon\varepsilon_0}\left(\frac{I}{Aq\mu_0E2\pi r^2}\right)^{\frac{1}{l}}. \qquad \text{(S 9.34)}$$

Making a change of variables, $r^2E=y$, after integration we obtain:

$$E=q^{\frac{l-1}{l+1}}\left(\frac{l+1}{l}\frac{1}{3\varepsilon\varepsilon_0}\right)^{\frac{l}{l+1}}\left(\frac{I}{A\mu 2\pi}\right)^{\frac{1}{l+1}}\frac{\left(r^3-r_0^3\right)^{\frac{l}{l+1}}}{r^2}. \qquad \text{(S 9.35)}$$

The constant of integration $r_0^{\;3}$ is calculated according to Ref. [49] using the boundary condition $E(r_0)=0$, where $r_0$ is the probe (cathode) radius. (In numerical estimates, we equal it to the contact radius, $r_0=a$). Substituting Eq. (S 9.35) into Eq. (S 9.33), we can obtain the injected charge distribution (at $\rho_f \ll \rho_t$, $\rho_i \approx \rho_t$):

$$\rho_i=\left(\frac{l}{l+1}\frac{3\varepsilon\varepsilon_0 I}{2\pi Aq^2\mu_0}\frac{1}{r^3-r_0^3}\right)^{\frac{1}{l+1}}, \qquad \text{(S 9.36)}$$

Using the regional approximation, we can estimate the radius $r_x$ for the SCLC region (Fig. S 9.3. (d)) from Eq. (S 9.36) at $\rho_i(r_x)=\rho_0$:

$$r_x = \left( \frac{l}{l+1} \frac{3\varepsilon\varepsilon_0 I}{2\pi A q^2 \mu_0} \rho_0^{-(l+1)} + r_0^3 \right)^{\frac{1}{3}}. \quad \text{(S 9.37)}$$

Estimation of $r_x$ at $r_0$=5 nm yields:

$$r_x = \left( \frac{1}{2} \frac{3\varepsilon\varepsilon_0 I}{2\pi A q^2 \mu_0 \rho_0^{2}} + r_0^3 \right)^{\frac{1}{3}} = \left( \frac{1.34 \cdot \left(10^{-28} - 10^{-26}\right)}{A} + 1.25 \cdot 10^{-25} \right)^{\frac{1}{3}} \text{m. At}$$

$I \geq \frac{l+1}{l} \frac{2\pi A q^2 \mu_0 \rho_0^{2} r_0^{3}}{3\varepsilon\varepsilon_0}$ ~ 1– 100 nA, $r_x \geq \sqrt[3]{2} \cdot r_0$, and, hence, $r_0$ can be neglected in Eqs. (S 9.37) and (S 9.35). Separating the SCLC and ohmic regions like in the 1D case (Fig. S 9.2), we can express the voltage drop across the SCLC region using Eq. (S 9.35). Using the boundary conditions $U - U_{rx} = \int_{r_0}^{r_x} E dr$ and neglecting $r_0$, after integration, we obtain:

$$U - U_{rx} = q^{\frac{l-1}{l+1}} \left( \frac{l+1}{l} \frac{1}{3\varepsilon\varepsilon_0} \right)^{\frac{l}{l+1}} \left( \frac{I}{A\mu_0 2\pi} \right)^{\frac{1}{l+1}} \left( r_x^{\frac{2l-1}{l+1}} - r_0^{\frac{2l-1}{l+1}} \right) \frac{l+1}{2l-1}. \quad \text{(S 9.38)}$$

Substituting $r_x$, Eq. (S 8.37), into (S 8.38), at high injection currents when $r_x > r_0$ we obtain an asymptotic solution for the I-V dependence:

$$U - U_{rx} \approx \left( \frac{l+1}{l} \frac{1}{3q\varepsilon\varepsilon_0} \right)^{\frac{1}{3}} \left( \frac{I}{A\mu_0 2\pi} \right)^{\frac{2}{3}} \frac{l+1}{2l-1} \rho_0^{\frac{1-2l}{3}}. \quad \text{(S 9.39)}$$

Using a relation for the spreading resistance in the semispherical case $R_s = \frac{1}{2\pi r_x \sigma_0}$, neglecting $r_0$ at $r_x > r_0$ for the I-V dependence, $U(I) = U - U_{rx} + \frac{I}{2\pi r_x \sigma_0}$, we obtain:

$$U(I) = \left( \frac{l+1}{2l-1} \frac{\rho_0^{1-l}}{A} + 1 \right) \left( \frac{(l+1)}{l} \frac{A\rho_0^{l}}{3\varepsilon\varepsilon_0} \right)^{\frac{1}{3}} \left( \frac{I}{2\pi\sigma_0} \right)^{\frac{2}{3}}. \quad \text{(S 9.40)}$$

The final relation Eq. (S 9.40) is a 3/2 power law: $I \sim U^{\frac{3}{2}}$. The law is valid for a trap-free solid ($l$ = 1) as well as for exponential and Gaussian trap distributions over energy. The relation can be used for description of asymptotic behavior of the 3D SCLC voltage-current dependence, $U_t - U_{rx}$, in the region between $t$ and $r_x$ (Fig. S 9.3 (d)).

Supplementary 10.

## Point contact resistance in terms of the Li-ion redistribution at positive and negative polarities on the tip

The calculations will be performed for the ohmic conduction regime in a uniform media. The balance between diffusion and migration fluxes of the Li-ions in the steady-state is described by a relation:

$$n(r) = n_0 \pm \frac{I}{\mu_{e^-} k_B T 2\pi r}. \tag{S 10.1}$$

Where $n_0$ is original Li-ion concentration in the material. Due to stoichiometry of the material, the concentration is limited by maximal $n_{\max}$ and minimal $n_{\min}$ values. Corresponding radii obtained from Eq. (S 10.1) for $n_{\min}$ and $n_{\max}$ are:

$$r_+ = \frac{I}{\mu_{e^-} k_B T 2\pi (n_0 - n_{\min})}, \tag{S 10.2}$$

and

$$r_- = \frac{I}{\mu_{e^-} k_B T 2\pi (n_{\max} - n_0)}. \tag{S 10.3}$$

They will be used in the following as integration limits.

Using the relation for concentration Eq. (S 10.1), the conductivity can be expressed as $\sigma(r) = \frac{1}{q\mu_{e^-}\left(n_0 + \frac{I}{\mu_{e^-} k_B T 2\pi r}\right)}$. The resistance, $R = \frac{1}{2\pi}\int_{r_0}^{\infty} \frac{dr}{r^2 \sigma(r)}$, in the hemispherical case for positive and negative polarity on the tip at $r_0 < r_+$ and $r_0 < r_-$ can be calculated according to the schematic representation of concentration distributions in Fig. S 10.1 by separation of integration over the regions before and after $r_+$ and $r_-$ (Fig. S 10(a), (b)) for the positive and negative polarity, respectively.

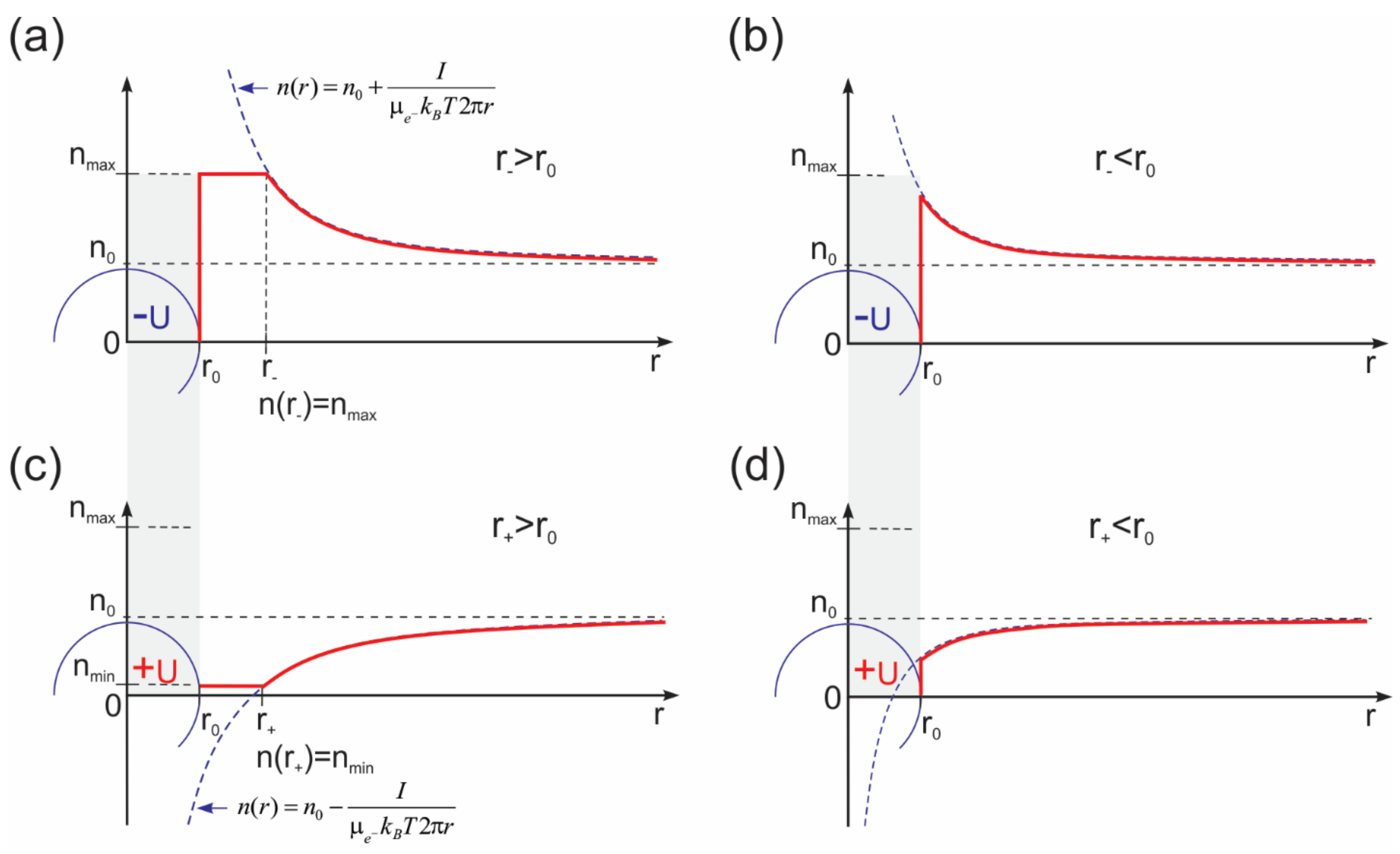

Figure S 10.1. Schematic profiles (red lines) of Li-ion concentration distributions under the tip radius of $r_0$ at equilibrium: (a), (b) for negative voltage on the tip; (c), (d) for positive voltage on the tip. (a), (c) Distributions at $r_0 < r_+$ and $r_0 < r_-$, accounting for the steric effect, which limits the maximal concentration to $n_{\max}$; the minimal concentration $n_{\min} \geq 0$. (b), (d) Distributions at $r_0 > r_+$, $r_0 > r_-$.

1. For positive polarity (Fig S 10.1 (c)).

$$R_b^+ = \frac{1}{2\pi}\int_{r_0}^{r_+}\frac{dr}{r^2\sigma_{\min}} + \frac{1}{2\pi}\int_{r_+}^{\infty}\frac{dr}{r^2\sigma(r)} = \frac{1}{2\pi q\mu_{e^-}}\left(\frac{1}{an_{\min}} - \frac{1}{r_+ n_{\min}} + \frac{1}{n_0}\int_{r_+}^{\infty}\frac{dr}{r^2 - \frac{\mu_{Li} I}{\mu_{e^-} 2\pi q D n_0} r}\right) = .$$

$$= \frac{1}{2\pi a\sigma_{\min}} - \frac{D}{I\cdot\mu_{Li}}\frac{(n_0 - n_{\min})}{n_{\min}} - \frac{D}{I\cdot\mu_{Li}}\ln\left(\frac{n_{\min}}{n_0}\right) = R_{\max} - \frac{D}{I\cdot\mu_{Li}}\left(\frac{n_0}{n_{\min}} - \ln\left(\frac{n_0}{n_{\min}}\right) - 1\right)$$

2. For negative polarity (Fig S 9.1 (a)).

$$R_b^- = \frac{1}{2\pi}\int_{r_0}^{r_-}\frac{dr}{r^2\sigma_{\max}} + \frac{1}{2\pi}\int_{r_-}^{\infty}\frac{dr}{r^2\sigma_0(r)} = \frac{1}{2\pi q\mu_{e^-}}\left(\frac{1}{an_{\max}} - \frac{1}{r_- n_{\max}} + \frac{1}{n_0}\int_{r_-}^{\infty}\frac{dr}{r^2 + \frac{\mu_{Li} I}{\mu_{e^-} 2\pi q D n_0} r}\right) = .$$

$$= \frac{1}{2\pi a\sigma_{\max}} - \frac{D}{I\cdot\mu_{Li}}\frac{n_{\max} - n_0}{n_{\max}} + \frac{D}{I\cdot\mu_{Li}}\ln\left(\frac{n_{\max}}{n_0}\right) = R_{\min} + \frac{D}{I\cdot\mu_{Li}}\left(\frac{n_0}{n_{\max}} - \ln\left(\frac{n_0}{\tilde{n}_{\max}}\right) - 1\right)$$

Using the Einstein relation $D = \frac{\mu_{Li} k_B T}{q}$, where $q$ is the ionic charge, finally we obtain:

$$R_b^+ = R_{\max} - \frac{k_B T}{I \cdot q}\left(\frac{n_0}{n_{\min}} - \ln\left(\frac{n_0}{n_{\min}}\right) - 1\right), \qquad \text{(S 10.4)}$$

$$R_b^- = R_{\min} + \frac{k_B T}{I \cdot q}\left(\frac{n_0}{n_{\max}} - \ln\left(\frac{n_0}{n_{\max}}\right) - 1\right). \qquad \text{(S 10.5)}$$

And at $r_0 > r_+$, $r_0 > r_-$:

$$R_b^+ = \frac{1}{2\pi}\int_{r_0}^{\infty}\frac{dr}{r^2\sigma^+(r)} = \frac{1}{2\pi q\mu_{e^-}}\frac{1}{n_0}\int_{r_0}^{\infty}\frac{dr}{r^2 - \frac{\mu_{Li} I}{\mu_{e^-} 2\pi q D n_0} r} = \frac{D}{\mu_{Li} I}\ln\left(\frac{1}{1 - \frac{\mu_{Li} I}{2\pi\sigma_0 D r_0}}\right),$$

$$R_b^- = \frac{1}{2\pi}\int_{r_0}^{\infty}\frac{dr}{r^2\sigma^-(r)} = \frac{1}{2\pi q\mu_{e^-}}\frac{1}{n_0}\int_{r_0}^{\infty}\frac{dr}{r^2 + \frac{\mu_{Li} I}{\mu_{e^-} 2\pi q D n_0} r} = \frac{D}{\mu_{Li} I}\ln\left(1 + \frac{\mu_{Li} I}{2\pi\sigma_0 D r_0}\right).$$

Where conductivity $\sigma_0 = q\mu_{e^-} n_0$, minimal and maximal resistance in Eqs. (S 10.4) and (S 10.5) are $R_{\min} = \frac{1}{2\pi r_0 q\mu_{e^-} n_{\max}}$ and $R_{\max} = \frac{1}{2\pi r_0 q\mu_{e^-} n_{\min}}$, respectively. Using an approximation for $r_+$ (Eq. (S 10.2)) at $n_{\min} \approx 0$: $r_+ \approx \frac{I}{2\pi\mu_{e^-} k_B T n_0}$, and the Einstein relation, we finally obtain:

$$R_b^+ = \frac{k_B T}{qI}\ln\left(\frac{1}{1 - \frac{r_+}{r_0}}\right) \approx \frac{k_B T}{qI}\frac{r_+}{r_0} + \frac{k_B T}{2qI}\left(\frac{r_+}{r_0}\right)^2 = \frac{1}{2\pi\mu_{e^-} q n_0 r_0}\left(1 + \frac{I}{4\pi\mu_{e^-} k_B T n_0 r_0}\right), \qquad \text{(S 10.6)}$$

$$R_b^- = \frac{k_B T}{qI}\ln\left(1 + \frac{r_+}{r_0}\right) \approx \frac{k_B T}{qI}\frac{r_+}{r_0} - \frac{k_B T}{2qI}\left(\frac{r_+}{r_0}\right)^2 = \frac{1}{2\pi\mu_{e^-} q n_0 r_0}\left(1 - \frac{I}{4\pi\mu_{e^-} k_B T n_0 r_0}\right). \qquad \text{(S 10.7)}$$

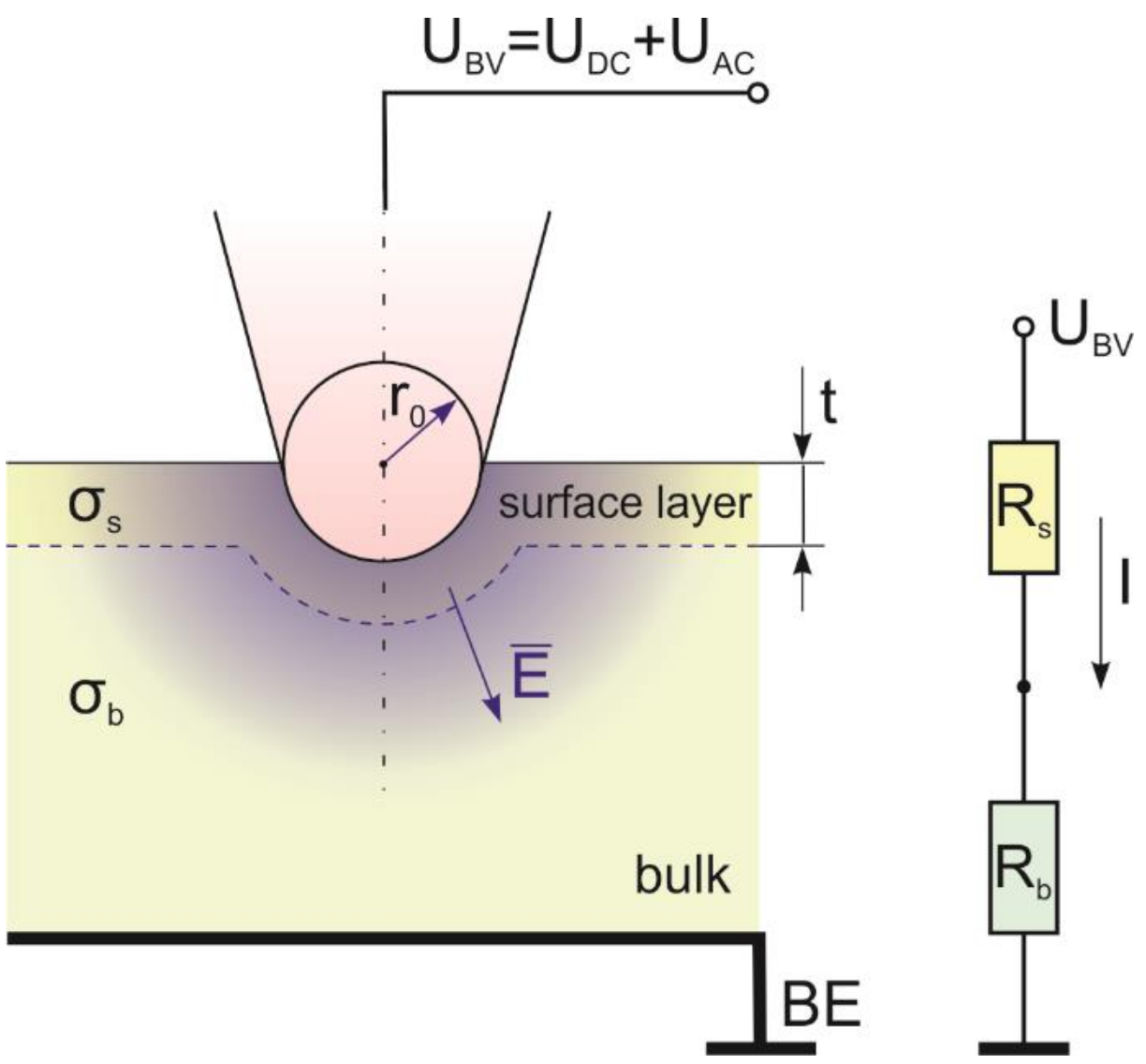

Figure. S 10.2. Schematic of the AFM probe contact with the LMO particle. Spherically symmetrical geometry is used for the calculation of the potential distribution in half-space near the tip. An equivalent electric circuit (right) includes: $R_s$ is the resistance of the surface layer of a thickness *t*, including the interface resistance between the metallic tip and LMO surface, $R_b$ is the spreading resistance of the bulk.

According to the model in Fig S 10.2, the bias voltage is the sum of the voltage drop across the surface layer and the voltage drop through the bulk: $U = U_s + U_b$. The ohmic conduction takes place in the bulk. Therefore, $r_0$ in the equations should be replaced by the sum $r_0 + t$, where $t$ is the thickness of the surface layer (Fig S 10.2). With the ohmic conduction in the bulk and the SCLC conduction in the surface layer, the total voltage drop is:

$$U(I) = I^{\frac{1}{\alpha}} \tilde{R}_s + I\tilde{R}_b, \qquad \text{(S 10.8)}$$

and the impedance is $R(I) = \frac{\tilde{R}_i^{\frac{1}{\alpha}}}{I^{\frac{\alpha-1}{\alpha}}} + \tilde{R}_S$. Substituting Eqs. (S 10.4) and (S 10.5) for $R_b$ in Eq. (S 10.8), we obtain:

$$U_+(I) = (I\rho_i)^{\frac{1}{\alpha}} + IR_{\max} - \frac{k_B T}{q}\left(\frac{n_0}{n_{\min}} - \ln\left(\frac{n_0}{n_{\min}}\right) - 1\right), \qquad \text{(S 10.9)}$$

$$U_-(I) = (I\rho_i)^{\frac{1}{\alpha}} + IR_{\min} + \frac{k_B T}{q}\left(\frac{n_0}{n_{\max}} - \ln\left(\frac{n_0}{n_{\max}}\right) - 1\right). \qquad \text{(S 10.10)}$$

In the case of a lithium ion, the ion charge $q_{Li} = |e|$, and at room temperature, $\frac{D}{\mu_{Li}} = \frac{k_B T}{q} = 0.026$ V. In turn, with Eqs. (S 10.6) and (S 10.7), we obtain:

$$U_+(I) = I^{\frac{1}{\alpha_+}} R_s + \frac{k_B T}{q} \ln\left( \frac{1}{1 - \frac{r_+}{r_0 + t}} \right), \tag{S 10.11}$$

$$U_-(I) = I^{\frac{1}{\alpha_-}} R_s + \frac{k_B T}{q} \ln\left( 1 + \frac{r_+}{r_0 + t} \right), \tag{S 10.12}$$

---